\documentclass{article}
\usepackage{psfig}
\usepackage{rotating}
\usepackage{emulateapj}
\usepackage{apjfonts}

\newcommand{\proptwid}{\mathrel{\raise.3ex\hbox{$\propto$}\mkern-14mu
             \lower0.6ex\hbox{$\sim$}}}

\makeatletter
\let\@internalcite\cite
\def\cite{\def\astroncite##1##2{##1\ ##2}\@internalcite}
\def\citey{\def\astroncite##1##2{##1\ (##2)}\@internalcite}
\def\@citex[#1]#2{\if@filesw\immediate\write\@auxout{\string\citation{#2}}\fi
  \def\@citea{}\@cite{\@for\@citeb:=#2\do
    {\@citea\def\@citea{; }\@ifundefined
       {b@\@citeb}{{\bf ??}\@warning
       {Citation `\@citeb' on page \thepage \space undefined}}%
{\csname b@\@citeb\endcsname}}}{#1}}
\def\@cite#1#2{#1\if@tempswa #2\fi}
\def\@biblabel#1{}
\def\astroncite#1#2{#1\ #2}
\makeatother

\def\errtwo#1#2#3{ $#1^{+ #2}_{- #3}$ }
\newcommand{\errtwa}[3]{#1^{+#2}_{-#3}}

\def\flux{{\rm ergs~cm^{-2}~s^{-1}}}
\def\pca{{PCA}}
\def\hexte{{HEXTE}}

\def\aproxgt{\mathrel{%
      \rlap{\raise 0.511ex \hbox{$>$}}{\lower 0.511ex \hbox{$\sim$}}}}
\def\aproxlt{\mathrel{%
      \rlap{\raise 0.511ex \hbox{$<$}}{\lower 0.511ex \hbox{$\sim$}}}}

\begin{document}

\slugcomment{Accepted to The Astrophysical Journal, April 6, 1999}

\lefthead{Wilms et al.}
\righthead{Low Luminosity States of the Black Hole Candidate GX~339$-$4. I.}

\title{Low Luminosity States of the Black Hole Candidate GX~339$-$4. I. \\ 
     ASCA and  Simultaneous Radio/RXTE Observations}

\author{J\"orn Wilms\altaffilmark{1}, Michael A. Nowak\altaffilmark{2}, 
James B. Dove\altaffilmark{2,3}, Robert P. Fender\altaffilmark{4}, Tiziana
Di~Matteo\altaffilmark{5,6}} 

\altaffiltext{1}{Institut f\"ur Astronomie und Astrophysik,
  Abt.~Astronomie, Waldh\"auser Str. 64, D-72076 T\"ubingen, Germany;
  wilms@astro.uni-tuebingen.de} 
\altaffiltext{2}{JILA, University of
  Colorado, Campus Box 440, Boulder, CO~80309-0440, USA;
  mnowak@rocinante.colorado.edu}  
\altaffiltext{3}{current address, CASA, University of
  Colorado, Campus 389, Boulder, CO~80309-0389, USA; dove@casa.colorado.edu}
\altaffiltext{4}{Astronomical Institute `Anton Pannekoek', University of
  Amsterdam, and Center for High Energy Astrophysics, Kruislaan 403, 1098
  SJ, Amsterdam, The Netherlands; rpf@astro.uva.nl}
\altaffiltext{5}{Institute of Astronomy, Madingley Road, Cambridge CB3 OHA,
  UK; tiziana@ast.cam.ac.uk}
\altaffiltext{6}{AXAF Fellow; current address: Center for Astrophysics,
  Cambridge, MA 01238, USA; tdimatteo@cfa.harvard.edu}

\received{1998 October 12}
\accepted{1999 April 6}

\begin{abstract}
  We discuss a series of observations of the black hole candidate
  GX~339$-$4 in low luminosity, spectrally hard states.  We present
  spectral analysis of three separate archival Advanced Satellite for
  Cosmology and Astrophysics (ASCA) data sets and eight separate Rossi
  X-ray Timing Explorer (RXTE) data sets.  Three of the RXTE observations
  were strictly simultaneous with 843\,MHz and 8.3--9.1\,GHz radio
  observations.  \emph{All} of these observations have (3--9\,keV) flux
  $\aproxlt 10^{-9}~{\rm ergs~s^{-1}~cm^{-2}}$.  The ASCA data show
  evidence for an $\approx 6.4$\,keV Fe line with equivalent width $\approx
  40$\,eV, as well as evidence for a soft excess that is well-modeled by a
  power law plus a multicolor blackbody spectrum with peak temperature
  $\approx 150$--$200$\,eV.  The RXTE data sets also show evidence of an Fe
  line with equivalent widths $\approx 20$--$140$\,eV.  Reflection models
  show a hardening of the RXTE spectra with decreasing X-ray flux; however,
  these models do not exhibit evidence of a correlation between the photon
  index of the incident power law flux and the solid angle subtended by the
  reflector.  `Sphere+disk' Comptonization models and Advection Dominated
  Accretion Flow (ADAF) models also provide reasonable descriptions of the
  RXTE data.  The former models yield coronal temperatures in the range
  $20$--$50$\,keV and optical depths of $\tau \approx 3$.  The model fits
  to the X-ray data, however, do not simultaneously explain the observed
  radio properties.  The most likely source of the radio flux is
  synchrotron emission from an extended outflow of size greater than
  ${\cal O}(10^7~GM/c^2)$.
\end{abstract}


%Correct footnote-numbering

\setcounter{footnote}{0}

\section{Introduction}\label{sec:intro}
The galactic black hole candidate (BHC) GX~339$-$4 is unique among
persistent sources in that it shows a wide variety of spectral states and
transitions among these states.  In presumed order of increasing bolometric
luminosity, GX~339$-$4 exhibits states with hard, power-law spectra (`off
state', \cite{ilovaisky:86a}; `low state', \cite{grebenev:91a}); a soft
state with \emph{no} evidence of a power law tail (`high state';
\cite{grebenev:91a}); and a very bright, soft state with extended power-law
tail (`very high state'; \cite{miyamoto:91a}).  There also are apparently
times when the flux is high, but the spectrum is not as soft as the `high'
or `very high state'.  \citey{mendez:97a} refer to this as the
`intermediate state'.  We also note that there is some evidence of overlap
between the states.  The broad-band (GRANAT/SIGMA) hard state data
presented by \citey{grebenev:91a} apparently represents a more luminous
state than does the broad-band soft state data taken with the same
instrument. [\citey{miyamoto:95a} has suggested the possibility of
hysteresis in galactic BHC state transitions.]  Similarly diverse sets of
states have been observed in X-ray transients such as Nova Muscae
(\cite{kitamoto:92a,miyamoto:94a}); however, GX~339$-$4 is closer to being
a persistent source.

Although there have been a number of observations of GX~339$-$4 in the
near-infrared and optical
(\cite{doxsey:79a,motch:83a,motch:85a,steiman:90a,imamura:90a,cowley:91b}),
including detection of a 14.8\,hr periodicity in the optical
(\cite{callanan:92a}), there is no convincing mass function for the system.
In the optical, the system is faint, variable ($M_{\rm V} \approx
16$--$20$), and reddened ($A_{\rm V} = 3.5$). The physical source of the
optical emission is unknown. It has been hypothesized that it is
\emph{entirely} dominated by the accretion disk, as the optical flux is
apparently anticorrelated with the soft X-ray emission
(\cite{steiman:90a,imamura:90a}). These properties of the emission have
made it difficult to obtain a good distance measurement, with estimates
ranging from 1.3\,kpc (\cite{predhel:91a}) to 8\,kpc (\cite{grindlay:79a}),
with many authors choosing 4\,kpc
(\cite{doxsey:79a,cowley:87a}).  A careful study of these
distance estimates is presented by \citey{zdziarski:98a} who argue for a
distance of $\approx 4$\,kpc.

GX~339$-$4 also has been detected in the radio (\cite{sood:94a}), and
possibly even has exhibited extended emission (\cite{fender:97a}).  Within
the hard state, the radio spectrum is flat/inverted with a spectral index
of $\alpha=0.1$--0.2 (\cite{fender:97a,corbel:98a}), where the radio flux
density $S_\nu \propto \nu^{\alpha}$.  Furthermore, in this state the radio
flux is correlated with the X-ray and gamma-ray fluxes (\cite{hanni:98a}),
but the radio flux disappears as GX~339$-$4 transits to a higher X-ray
flux/softer state (Fender 1998, priv.\ comm.),  which is comparable to the
behavior of Cyg~X-1 (\cite{pooley:98a}).

During the Rossi X-ray Timing Explorer (RXTE) Cycle~2 observing phase
(1996~December -- 1998~February), we performed a series of eight RXTE 
observations of GX~339$-$4.  The first three observations were spaced a week
apart from one another from 1997 February~4 to 1997 February~18.  These
three observations were scheduled to be simultaneous with 8.3--9.1\,GHz
radio observations that were conducted at the Australian Telescope Compact
Array (ATCA).  The results of the radio observations have been reported by
\citey{corbel:98a}.  Additionally, three 843\,MHz observations performed at
the Molongolo Observatory Synthesis Telescope (MOST) and reported by
\citey{hanni:98a} are also simultaneous with these RXTE observations.  

This paper is structured as follows.  We discuss the spectral analysis of
archival Advanced Satellite for Cosmology and Astrophysics (ASCA) data in
section~\ref{sec:asca}.  We look for evidence of Fe lines in the data and
we characterize the soft ($\aproxlt 1$\,keV) X-ray data.  In
section~\ref{sec:xte} we present the RXTE data.  We first discuss the All
Sky Monitor (ASM) data, and then we discuss the pointed observations.  We
perform spectral analysis much akin to that which we considered for Cyg~X-1
(\cite{dove:98a}).  Here, however, we consider Advection Dominated
Accretion Flow (ADAF) models as well by using the models described by
\citey{dimatteo:98a}.  We discuss the implications of the simultaneous
radio data in section~\ref{sec:radio}.  In section~\ref{sec:discuss} we
discuss the implications of the X-ray observations for theoretical models.
We summarize our results in section~\ref{sec:summ}. We present timing
analysis of the RXTE data in a companion paper (\cite{nowak:98c},
henceforth paper~II).

\section{Archival ASCA Observations}\label{sec:asca}

\begin{table*}
\caption{\small Log of the ASCA observations. \label{tab:ascalog}}
\smallskip
\center{
\begin{tabular}{lllll}
\hline
\hline
\noalign{\vspace*{1mm}}
Obs.    & Date              & Integration time &  SIS0      & 3--9\,keV
        Flux \\ 
        &                   & (ksec)             & (cts s$^{-1}$)&
        ($10^{-9}~\flux$)\\
\noalign{\vspace*{1mm}}
\hline
\noalign{\vspace*{1mm}}
1       & 1994 August 24    & 15   &  3.6 & 0.11 \\
\noalign{\vspace*{1mm}}
2       & 1994 September 12 & 17   &  6.4 & 0.19 \\
\noalign{\vspace*{1mm}}
3       & 1995 September 8     & 30   & 17.5 & 0.63 \\
\noalign{\vspace*{1mm}}
\hline
\end{tabular}}
\tablecomments{All  observations were taken in Bright 1-CCD mode. SIS0:
  filtered SIS0 count rate.}
\end{table*}

The ASCA archives contain four observations of the GX~339$-$4 region.  A
log of the observations is given in in Table~\ref{tab:ascalog}.  In
appendix~\ref{sec:ascaapp}, we describe the methods that we used to
extract, filter, and analyze these ASCA observations. To the best of our
knowledge, an analysis of these observations has not been published
previously, except for a power spectrum for one of the observations (date
not given, \cite{dobrin:97a}).  The first of the observations (1993
September 16) did not detect the source, with the upper limit to the
3--9\,keV flux being $\approx 10^{-12}~{\rm ergs~s^{-1}~cm^{2}}$.  As we
will discuss further below, the inferred 3--9\,keV fluxes for the remaining
three observations (Table~\ref{tab:ascalog}) are lower by a factor of two
to ten than the fluxes of the RXTE observations discussed in
\S\ref{sec:xte}.

We chose to fit the ASCA data with a phenomenological model consisting of a
multicolor blackbody spectrum plus a broken power law, considered with and
without a narrow Gaussian line at $\approx 6.4$\,keV.  These fits are
similar to those performed for ASCA observations of the hard state of
Cyg~X-1 (\cite{ebisawa:96b}), which shows evidence for a weak and narrow Fe
line with equivalent width $\approx 40$\,eV, as well as for a soft excess
well-modeled as a multicolor blackbody with peak temperature $\approx
150$\,eV.

\begin{table*}
\caption{\small Parameters for a multicolor blackbody plus broken power law
  plus Gaussian line fits to ASCA data. \label{tab:ascafita}} 
\smallskip
{\small
\center{
\begin{tabular}{llllllllllll}
\hline
\hline
\noalign{\vspace*{1mm}}
Date & $T_{\rm in}$ & $A_{\rm dbb}$ & $\Gamma_1$ & $E_{\rm b}$ & $\Gamma_2$
& $A_{\rm bpl}$ & $E_{\rm l}$ & $A_{\rm l}$ & EW & $\chi^2$/dof &
$\chi^2_{\rm red}$ \\
& (keV) & ($\times 10^4)$ & & (keV) & & ($\times 10^{-2}$) & (keV) &
($\times 10^{-4}$) & (eV) \\ 
\noalign{\vspace*{1mm}}
\hline
\noalign{\vspace*{1mm}}
1994 Aug. 24 & \errtwo{0.14}{0.01}{0.02} & \errtwo{2.2}{0.9}{1.6} &
\errtwo{1.78}{0.03}{0.03} & \errtwo{3.4}{0.4}{0.5} &
\errtwo{1.62}{0.04}{0.03} & \errtwo{4.2}{0.1}{0.1} & \nodata & \nodata & \nodata &
1500/1439 & 1.04 \\
\noalign{\vspace*{1mm}}
1994 Aug. 24 & \errtwo{0.14}{0.01}{0.02} & \errtwo{2.2}{0.7}{1.6} &
\errtwo{1.78}{0.03}{0.03} & \errtwo{3.3}{0.5}{0.6} &
\errtwo{1.64}{0.03}{0.03} & \errtwo{4.2}{0.1}{0.1} & {\it 6.4} &
\errtwo{0.6}{0.4}{0.3} & \errtwo{34}{25}{19} & 1494/1438 & 1.04 \\
\noalign{\vspace*{1mm}}
1994 Sept. 12 & \errtwo{0.15}{0.01}{0.01} & \errtwo{2.5}{0.6}{1.0} &
\errtwo{1.81}{0.02}{0.01} & \errtwo{3.8}{0.2}{0.2} &
\errtwo{1.56}{0.03}{0.03} & \errtwo{7.2}{0.1}{0.1} & \nodata & \nodata & \nodata &
1603/1621 & 0.99 \\
\noalign{\vspace*{1mm}}
1994 Sept. 12 & \errtwo{0.15}{0.00}{0.00} & \errtwo{2.5}{0.1}{0.4} &
\errtwo{1.81}{0.01}{0.01} & \errtwo{3.8}{0.1}{0.1} &
\errtwo{1.59}{0.01}{0.02} & \errtwo{7.2}{0.0}{0.1} &
\errtwo{6.36}{0.08}{0.09} & \errtwo{1.6}{0.7}{0.2} & \errtwo{56}{26}{7} &
1580/1619 & 0.98 \\
\noalign{\vspace*{1mm}}
1995 Sept. 08 & \errtwo{0.19}{0.00}{0.01} & \errtwo{2.4}{0.1}{0.3} &
\errtwo{1.93}{0.02}{0.01} & \errtwo{3.7}{0.1}{0.1} &
\errtwo{1.60}{0.01}{0.02} & \errtwo{25.6}{0.3}{0.3} & \nodata & \nodata & \nodata &
2597/1838 & 1.41 \\
\noalign{\vspace*{1mm}}
1995 Sept. 08 & \errtwo{0.19}{0.00}{0.00} & \errtwo{2.4}{0.0}{0.1} &
\errtwo{1.93}{0.01}{0.00} & \errtwo{3.7}{0.1}{0.0} &
\errtwo{1.63}{0.01}{0.01} & \errtwo{25.6}{0.0}{0.1} &
\errtwo{6.51}{0.07}{0.06} & \errtwo{3.3}{0.6}{0.5} & \errtwo{40}{7}{6} &
2523/1836 & 1.37 \\
\noalign{\vspace*{1mm}}
\hline
\end{tabular} }

%%% Local Variables: 
%%% mode: latex
%%% TeX-master: t
%%% End: 

}
\tablecomments{$T_{\rm in}$: peak 
  multicolor blackbody temperature. $A_{\rm dbb}$: multicolor blackbody
  normalization. $\Gamma_1$, $\Gamma_2$: broken power law photon indices. 
  $E_{\rm b}$: break energy. $A_{\rm bpl}$: Power law normalization 
  (photons\,keV$^{-1}$\,cm$^{-2}$\,s$^{-1}$ at 1\,keV). $E_{\rm l}$: line
  energy.   $A_{\rm l}$: Line normalization
  (photons\,cm$^{-2}$\,s$^{-1}$ in the line).  EW: line equivalent width.
    Uncertainties are at the 90\% confidence level for one interesting
  parameter ($\Delta \chi^2 = 2.71$). }
\end{table*}

\begin{figure*}
\centerline{
\psfig{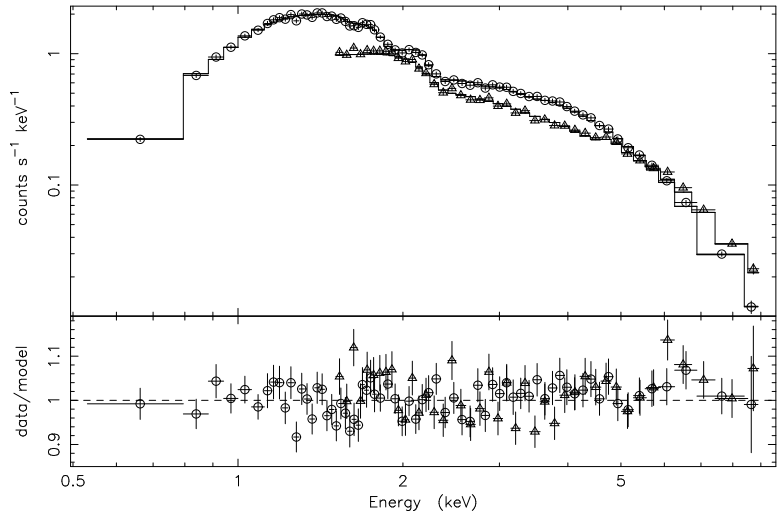} }
\caption{\small GX~339$-$4 ASCA observation of 1994 August~24 with 
  energy bins rebinned by a factor of 25 for clarity. 
  Model and associated residuals (data/model) are for the best
  fit multicolor blackbody plus broken power law \emph{without} a Gaussian
  line component.  For clarity, only the SIS0 (circles) and GIS2
  (triangles) data are shown.}
\label{fig:ascaphem}
\end{figure*}

The fits with the phenomenological models yield $\chi^2_{\rm red}$ ranging
from $0.98$ to $1.4$. The brightest data set showed the greatest evidence
for structure beyond this simple model.  A sample fit is shown in
Figure~\ref{fig:ascaphem}. Note that the neutral hydrogen column was fixed
to $6 \times 10^{21}~{\rm cm}^{-2}$.  Allowed to freely vary, the neutral
hydrogen column tended to float between $4$ and $8 \times 10^{21}~{\rm
  cm}^{-2}$, depending upon what combination of phenomenological models was
chosen, with minimal changes in the $\chi^2$ of the fits.  Associated with
these changes in best fit neutral hydrogen column were $\aproxgt \pm 30\%$
changes of the best fit peak temperature of the multicolor blackbody and
even larger changes (factors of $\approx 3$) in the best fit normalization
of the multicolor blackbody component.  We should thus associate
systematic error bars with these two parameters that are somewhat
larger than the statistical error bars presented in
Table~\ref{tab:ascafita}.

All fits improved with the addition of a narrow Gaussian line.  In all fits
we fixed the line width to $0.1$\,keV (see  \cite{ebisawa:96b}; who always
found $\sigma < 0.1$\,keV in fits to ASCA data of Cyg~X-1), and for the
lowest flux data set we also fix the line energy to 6.4\,keV. For the
lowest flux data set the $\Delta \chi^2 = 5.5$ for one additional
parameter.  By the F-test (\cite{bevington}), this is an improvement to the
fit for one additional parameter at the 98\% confidence level.  The other
two data sets show even more significant improvements to the $\chi^2$.  The
best fit line equivalent widths ranged from $\approx 30$ to 60\,eV.  There
is no compelling evidence for a strong flux dependence to the equivalent
width of the line.  

The transition to the bright, soft state typically occurs at 3--9\,keV
luminosities $\aproxgt 10^{-9}~{\rm ergs~cm^{-2}~s^{-1}}$; i.e., a factor
of two to ten brighter than these ASCA observations.  Thus these
observations offer useful tests of ADAF models, which are hypothesized to
be most relevant to low-luminosity, hard state systems
(\cite{narayan:96e,esin:97c}).  ADAF models predict a detectable
correlation between the temperature of the soft excess, the strength of the
Fe line, and the source luminosity.  They hypothesize that the luminosity
decay of BH transients is due, in part, to an increase of the radius at
which the accretion flow transits from cold, geometrically thin, and
radiatively efficient to hot, geometrically thick, and advective
(\cite{esin:97c}).  In some models, the transition radius can grow to as
large as ${\cal O}(10^4~R_{\rm G})$, where $R_{\rm G} \equiv GM/c^2$.
(Such large transition radii are \emph{not} a strict requirement of ADAF
models;  in \S\ref{sec:adaf} we show that somewhat smaller transition
radii, $\approx 200$--$400~R_{\rm G}$, are preferred for ADAF models of the
RXTE data.) As discussed by \citey{esin:97c}, one then expects the peak
temperature of the soft excess to decrease below $150$\,eV and the
equivalent width of any Fe line to decrease to values less than $\approx
30$\,eV.

The best fit equivalent widths found for GX~339$-$4 are greater than can be
accommodated in ADAF models with a large transition radius, and they are
also slightly larger than predicted by the `sphere+disk' corona models
described in \S\ref{sec:corona} (see also \cite{dove:97b,dove:98a}).  These
latter models have a similar geometry to the ADAF models, and they often
posit a coronal radius $\aproxlt 100~R_{\rm G}$.  Likewise, we do not
detect any large decreases in the best-fit disk temperatures with
decreasing luminosity.  Although it is dangerous to make a one-to-one
correspondence between a phenomenological fit component and a
\emph{physical} component, these best-fit values are suggestive of, but not
definitive proof of, temperatures hotter than can be accommodated in models
where cold, soft X-ray emitting material exists at very large radii.

\section{RXTE Observations}\label{sec:xte}
\subsection{The Monitoring Campaign}\label{sec:asm}

\begin{figure*}
\centerline{
\psfig{figure=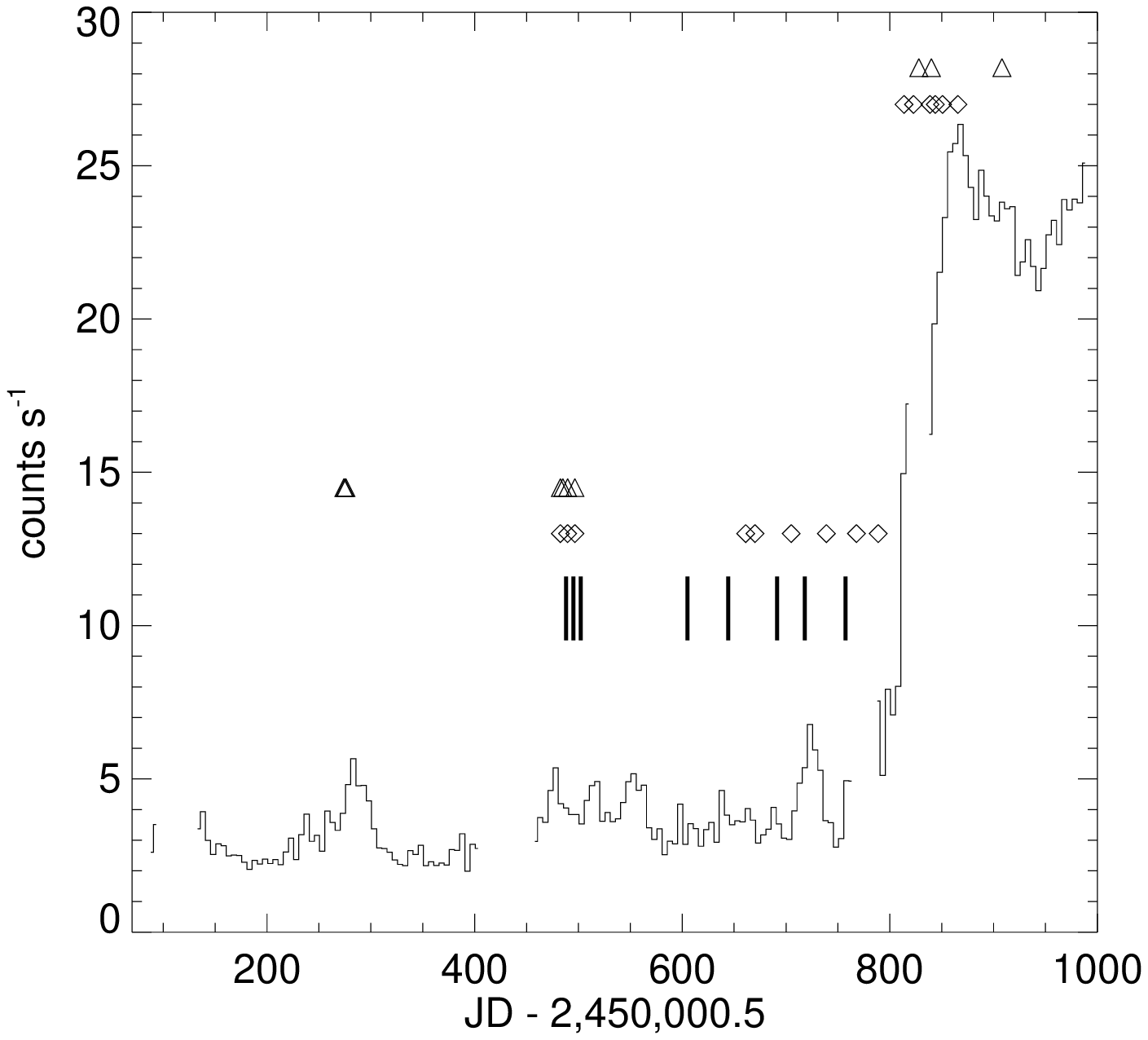,width=0.45\textwidth} 
\psfig{figure=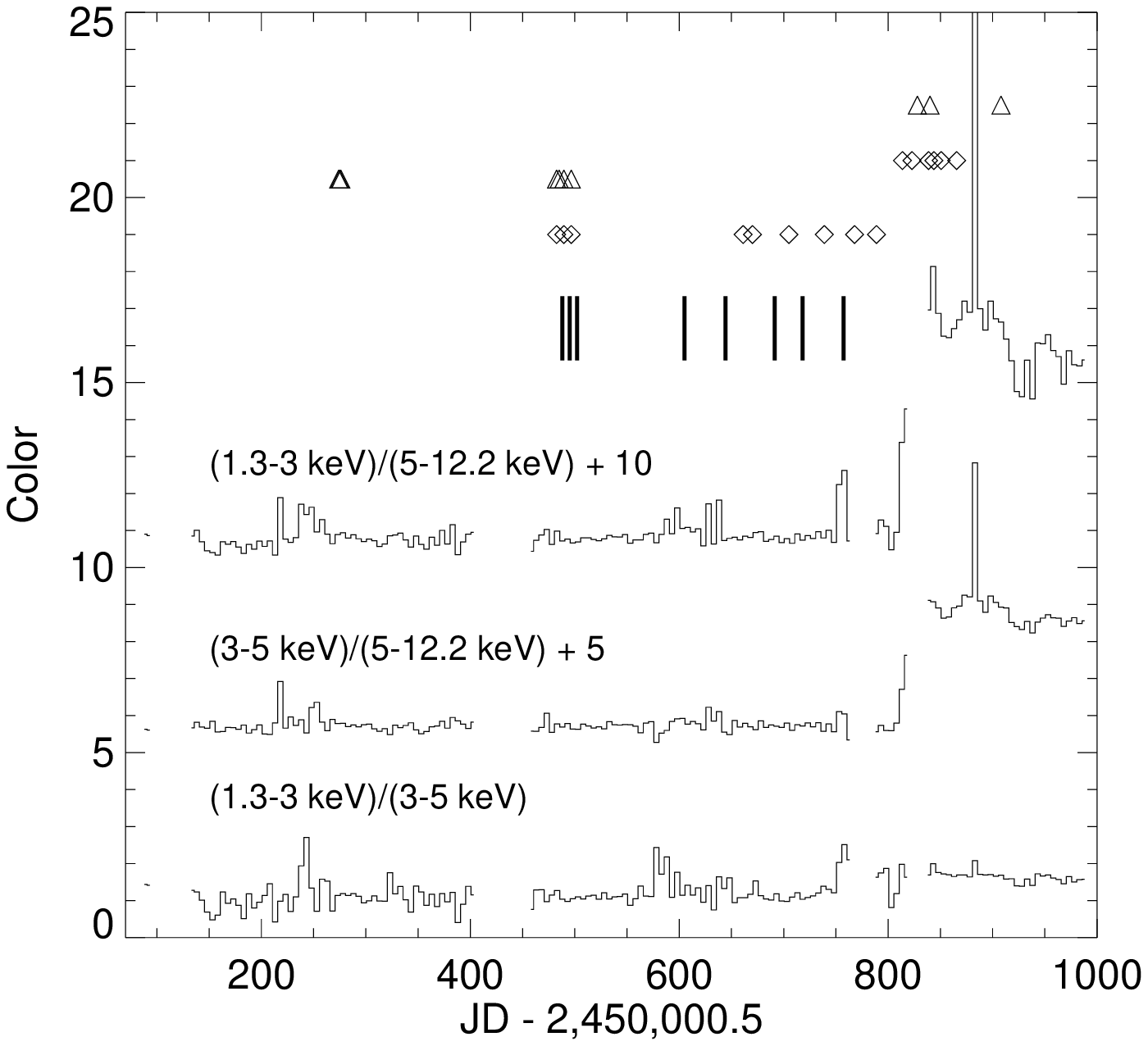,width=0.45\textwidth} }
\caption{\small \emph{Left:} RXTE All Sky Monitor data for GX~339$-$4 (5 day
  averages in the 1.3--12.2\,keV band) vs. Truncated Julian Date (TJD)
  $\equiv$ Julian Date (JD) $-2450000.5$.  Dashes indicate dates of our
  RXTE pointed observations, diamonds indicate dates of MOST radio
  observations (Hannikainnen et al. 1998), and triangles indicate dates of
  ATCA radio observations (Fender et al. 1997, Corbel et al.  1998).
  \emph{Right:} ASM colors for GX~339$-$4 vs TJD.  Colors shown are:
  (1.3--3\,keV)/(3-5\,keV), (3--5\,keV)/(5--12\,keV)+5, and
  (1.3--3\,keV)/(5--12.2\,keV)+10 (the latter two ratios have been offset
  for clarity).}
\label{fig:asm}
\end{figure*}

To study the long-term behavior of GX~339$-$4, and to place our pointed
observations within the context of the overall behavior of the source, we
used data from the All Sky Monitor (ASM) on RXTE. The ASM provides
lightcurves in three energy bands, 1.3--3.0\,keV, 3.0--5.0\,keV,
and 5.0--12.2\,keV, typically consisting of several 90\,s
measurements per day (see \cite{levine:96a,remillard:97a,lochner:97a}).
In Figure~\ref{fig:asm} we present the ASM data of GX~339$-$4 up until
Truncated Julian Date (TJD) $\approx 1000$ (1998 July 6).  We also indicate
the dates of our RXTE observations, as well as the dates of
ATCA and MOST radio observations (\cite{fender:97a,corbel:98a,hanni:98a}).
We discuss the long timescale variability of this lightcurve in paper~II.

\begin{table*}
\caption{\small Approximate expected ASM colors for the different states of
  GX~339$-$4. \label{tab:asmcol}}
\smallskip
\center{
\begin{tabular}{llll}
\hline
\hline
\noalign{\vspace*{1mm}}
State         & count rate & 1.3--3\,keV/ 
  & 1.3--3\,keV/ \\ 
          & cts\,s$^{-1}$ &  3--5\,keV  &  5--12.2\,keV                 \\
\noalign{\vspace*{1mm}}
\hline
\noalign{\vspace*{1mm}}
low       &   7   &  $2$  & $2$ \\
\noalign{\vspace*{1mm}}
high      &  15   &  $4$  & $30$\\
\noalign{\vspace*{1mm}}
very high &  60   &  $4$  & $15$\\
\noalign{\vspace*{1mm}}
\hline
\end{tabular}}
\end{table*}

Based upon model fits to the observations of \citey{grebenev:91a} (low and
high state), and \citey{miyamoto:91a} (very high state), we expect the
different states of GX~339$-$4 to have ASM count rates as indicated in
Table~\ref{tab:asmcol}.  The ASCA and RXTE observations discussed here are
most characteristic of weak to average luminosity hard states. Confirmation
that the eight RXTE observations taken between TJD 481 and 749 do indeed
represent a typical low/hard state comes from the broad band spectral
analysis presented in \S\ref{sec:xte}, as well as from the timing analysis
presented in paper II.  The X-ray variability of these observations show
root mean square variability of ${\cal O}(30\%)$ and show a power spectrum
(PSD) that, roughly, is flat below 0.1\,Hz, $\propto f^{-1}$ between
0.1--3\,Hz, and $\propto f^{-2}$ above 3\,Hz. Time lags and coherence
functions were also comparable to previously observed hard states of Cygnus
X-1 (see paper II, and references therein). Further discussion and analyses
of the timing data can be found in paper II.

The transition to a higher flux level that occurs at TJD $\approx 800$
appears to have a characteristic flux of a high state, but is not as soft
in the 2--10\,keV bands as expected from the above cited high and very high
states.  This might be an example of what \citey{mendez:97a} refer to as an
`intermediate state' between hard and soft.  No pointed RXTE observations
were taken during the transition, and four pointed observations, which were
not part of our monitoring campaign, occured shortly after the transition.
Detailed confirmation of the spectral state suggested by the ASM data
awaits analysis of these pointed observations.  Note that the radio flux
became quenched over the course of this transition to a higher ASM flux
level (\cite{hanni:98a}).

The variations observed in both the ASM lightcurve (prior to TJD $\approx
800$) and the pointed RXTE observations discussed below represent more than
a factor five variation in observed flux.  Comparable variations have been
observed in the radio, and furthermore the radio lightcurves show evidence
of a correlation with both the ASM and Burst and Transient Survey Explorer
(BATSE) lightcurves (\cite{hanni:98a}).

\subsection{PCA and HEXTE Observations}
In this section we present the results from our analysis of the data
from the two pointed instruments on RXTE: the Proportional Counter Array
(PCA), and the High Energy X-ray Timing Experiment (HEXTE). See
appendix~\ref{sec:rxteapp} for a description of the instruments and of the
details of the data extraction and processing. A log of the
RXTE pointed observations and the simultaneous radio observations is given
in Table~\ref{tab:log}. 

\begin{table*}
\caption{\small Log of RXTE and radio observations. \label{tab:log}}
\smallskip
\centerline{\small
\begin{tabular}{lllcccccccc}
\hline
\hline
\noalign{\vspace*{1mm}}
Obs & Date & TJD & $T$ & Rate & 3--9\,keV  & 9--30\,keV &
     30--100\,keV  & ATCA  & MOST &  $\alpha$ \\
    &  &    & (ksec) & (cps) & 
    & (${\rm 10^{-9}~erg~cm^{-2}~s^{-1}}$) 
    &  &  (mJy) 
    &  (mJy) \\
\noalign{\vspace*{1mm}}
\hline
\noalign{\vspace*{1mm}}
01  & 1997 Feb. 02  & 481 & 11 & 830  & 1.07  & 1.68 & 2.65  
    & $9.1\pm0.2$ & $7.0\pm0.7$ & 0.11 \\
\noalign{\vspace*{1mm}}
02  & 1997 Feb. 10  & 489 & 10 & 730  & 0.94  & 1.50 & 2.41 
    & $8.2\pm0.2$ & $6.3\pm0.7$ & 0.11 \\
\noalign{\vspace*{1mm}}
03  & 1997 Feb. 17  & 496 &  8 & 700  & 0.90  & 1.43 & 2.35  
    & $8.7\pm0.2$ & $6.1\pm0.7$ & 0.15 \\
\noalign{\vspace*{1mm}}
04  & 1997 Apr. 29  & 567 & 10 & 470  & 0.60  & 0.97 & 1.55  \\
\noalign{\vspace*{1mm}}
05  & 1997 Jul. 07  & 636 & 10 & 200  & 0.25  & 0.43 & 0.75  \\
\noalign{\vspace*{1mm}}
06  & 1997 Aug. 23  & 683 & 11 & 650  & 0.74  & 1.18 & 1.98  \\
\noalign{\vspace*{1mm}}
07  & 1997 Sep. 19  & 710 & 10 & 730  & 0.96  & 1.48 & 2.36  \\
\noalign{\vspace*{1mm}}
08  & 1997 Oct. 28  & 749 & 10 & 480  & 0.63  & 1.01 & 1.68  \\
\noalign{\vspace*{1mm}}
\hline
\end{tabular}}
\tablecomments{We list: $T$, the
  duration of the RXTE observations; the average PCA count rate; the
  average (3--9\,keV), (9--30\,keV), and (30--100\,keV) energy fluxes (all
  normalized to the PCA calibration); the flux density of the 8.3--9.1\,GHz
  ATCA observations; the flux density of the 843\,MHz MOST observations;
  and $\alpha = \Delta \log \nu / \Delta \log S_\nu$, the spectral index of
  the radio observations 
  (\protect\cite{fender:97a,corbel:98a,hanni:98a}).} 
\end{table*}

As we show in appendix~\ref{sec:rxteapp}, there is a difference in the
power-law slopes obtained from an analysis of spectra of the Crab with both
instruments upon RXTE.  In order to minimize the impact of this difference
in the instrumental calibration onto the data analysis, we primarily
analyze the data from both instruments individually and use the difference
in model parameters between instruments as a gauge of the systematic
uncertainties. We do perform some joint analysis of PCA and HEXTE data
using various reflection models.  In the following sections we discuss in
detail the implications of the calibration uncertainty for our analysis.

\begin{figure*}
\centerline{
\psfig{figure=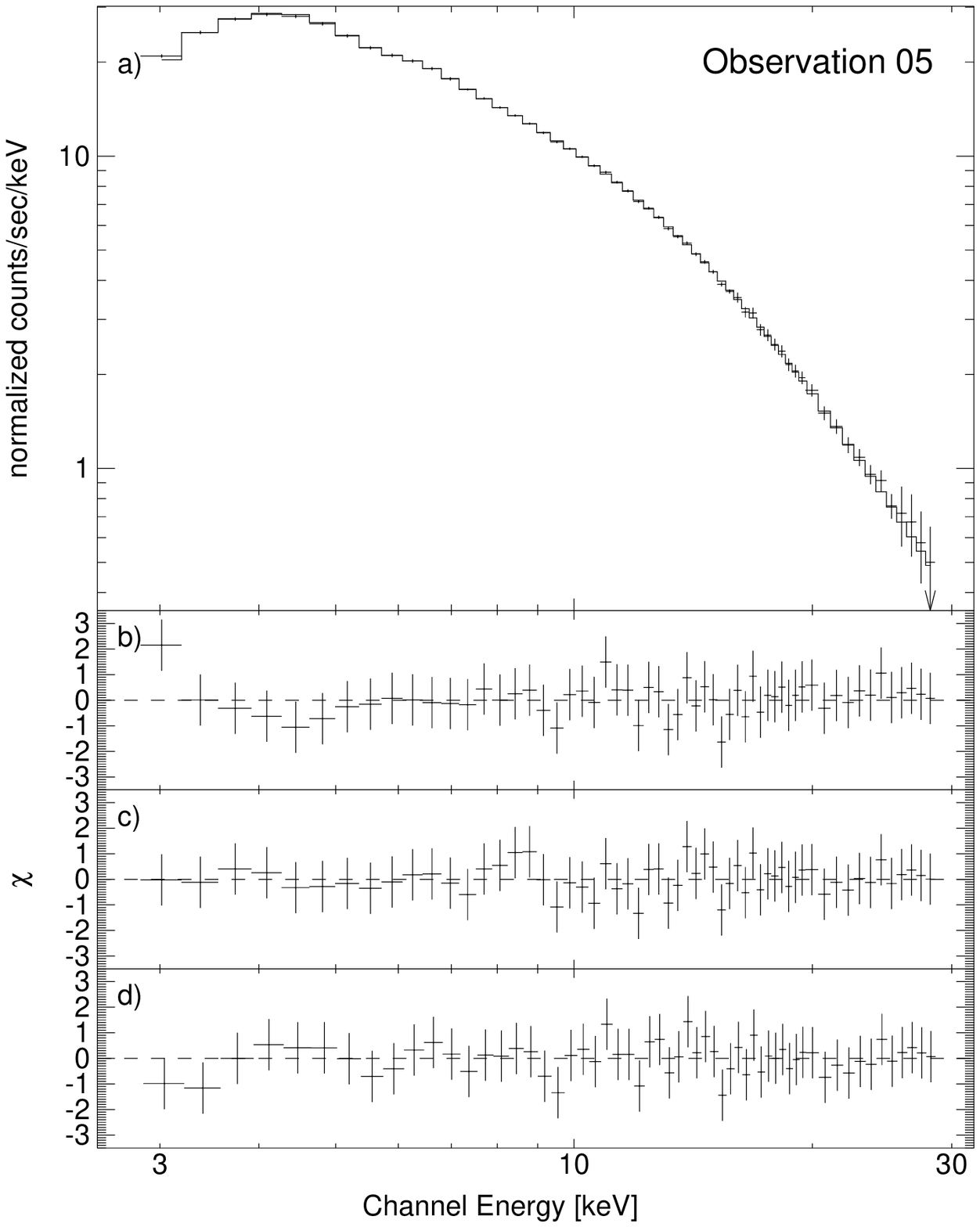,width=0.45\textwidth}
\psfig{figure=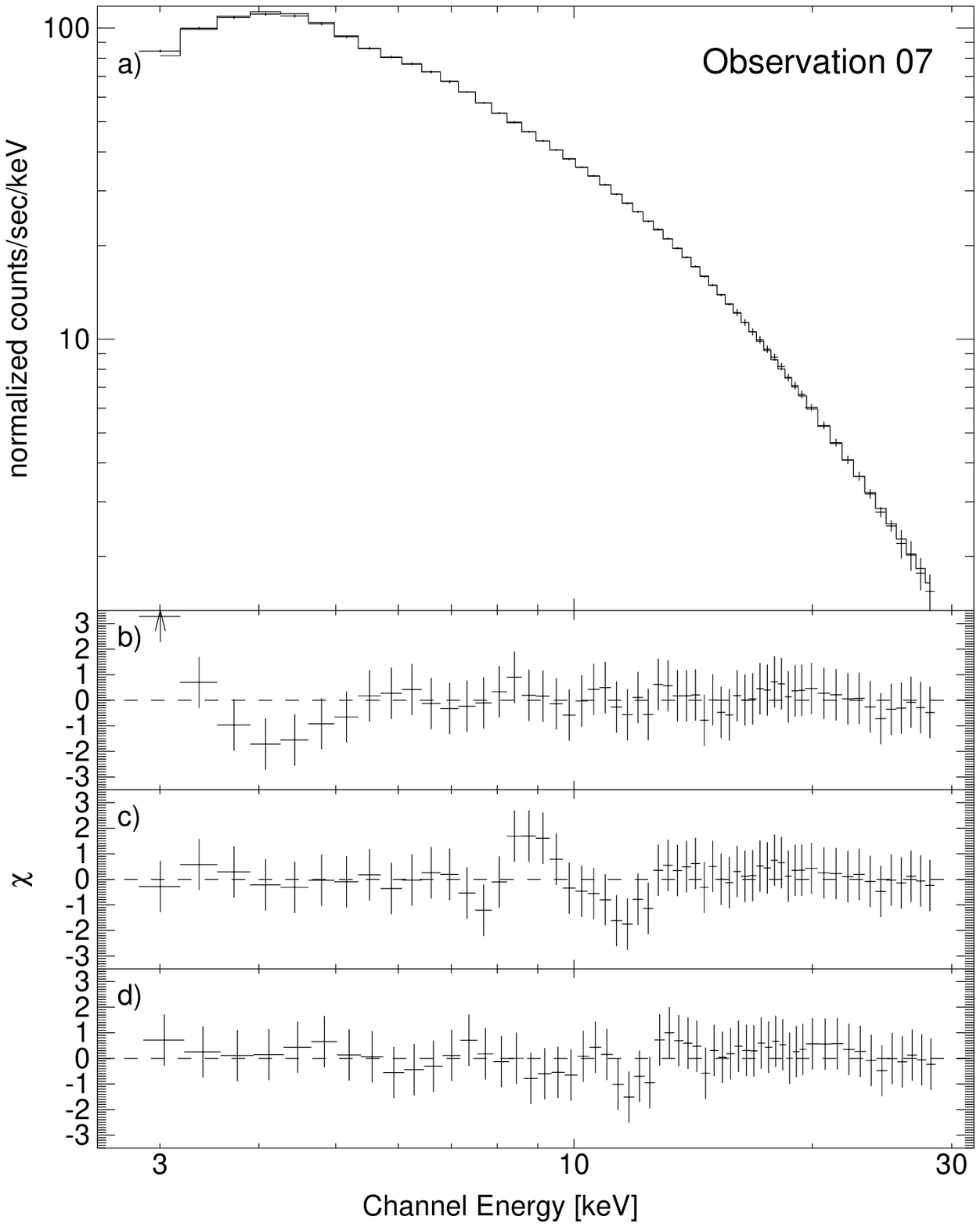,width=0.45\textwidth}}
\caption{\small Spectral modeling of the PCA data from Observation~5
  (left) and~7 (right). Residues are shown as the contribution to $\chi$.
  a) Count rate spectrum and the best-fit broken power-law with Gaussian
  line, b) Contribution to $\chi$ from the broken power-law with Gaussian
  line, c) Contribution to $\chi$ from the ionized reflection model
  (pexriv) with Gaussian line, d) Contribution to $\chi$ from the
  sphere+disk model with additional Gaussian line.}
\label{fig:pcaalone}
\end{figure*}

\begin{figure*}
\centerline{
\psfig{figure=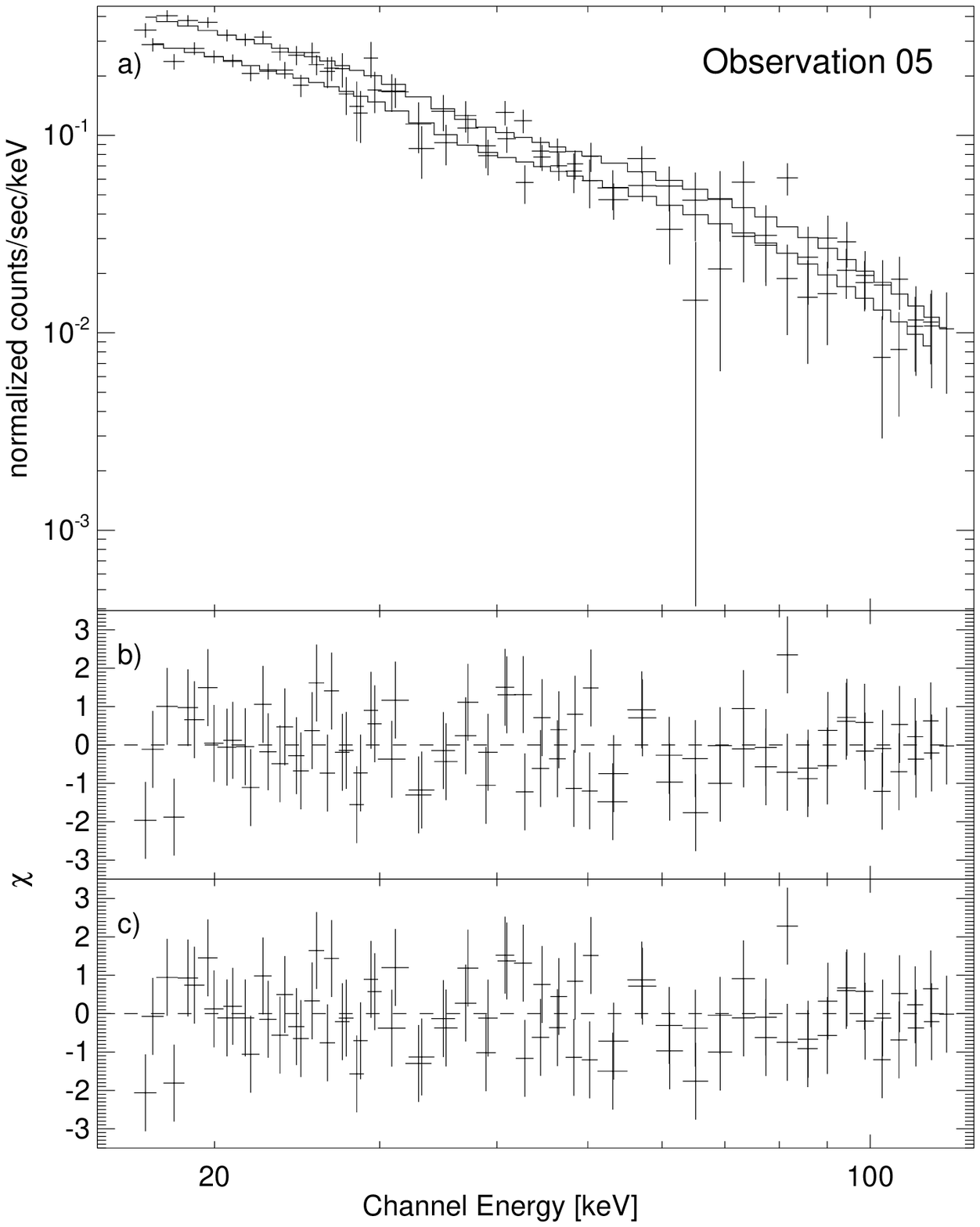,width=0.45\textwidth}
\psfig{figure=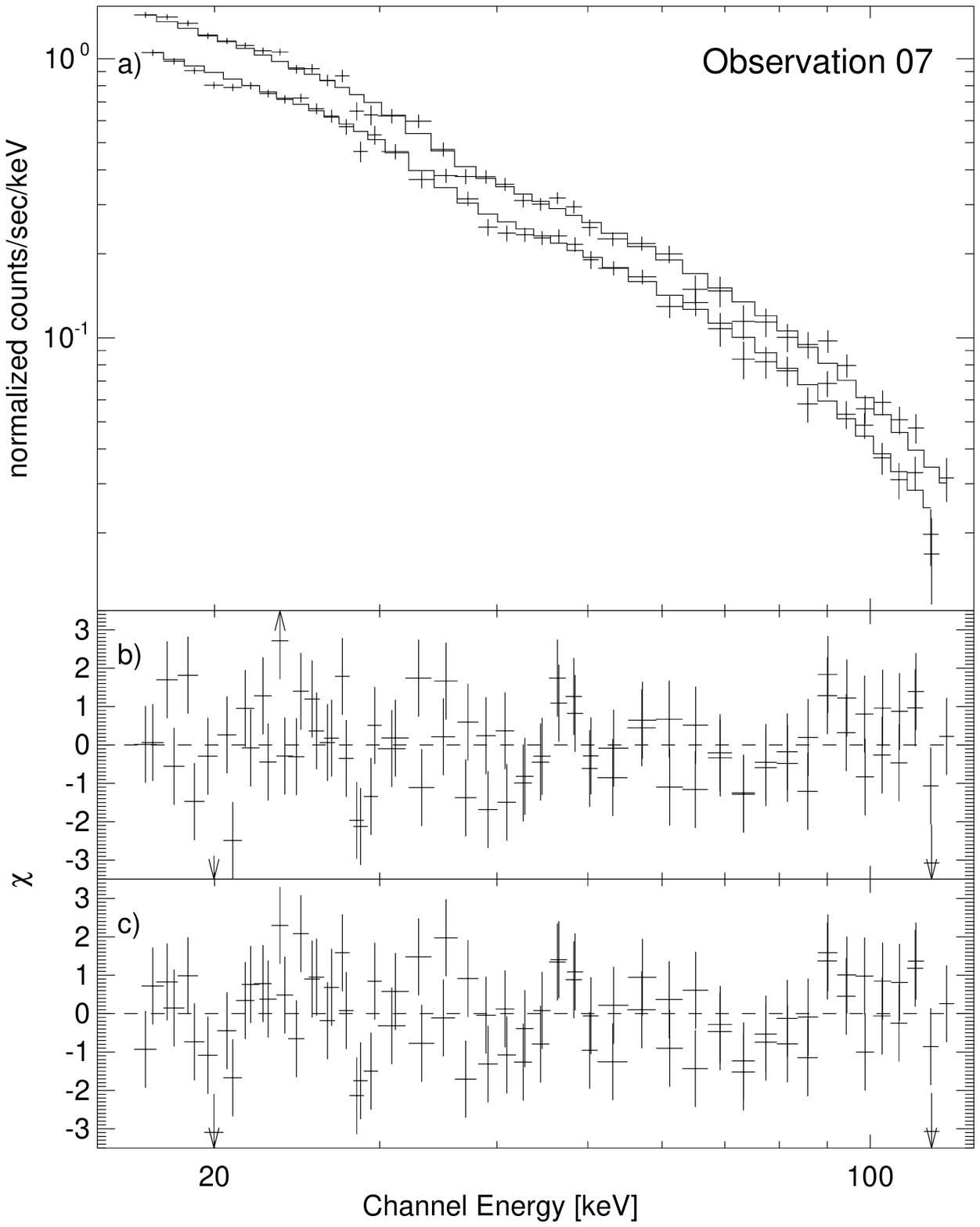,width=0.45\textwidth}}
\caption{\small Spectral modeling of the HEXTE data from observation~5
  (left) and~7 (right). Residues are shown as the contribution to
  $\chi$. a) Count rate spectrum and the best-fit power-law with
  exponential cutoff, b) Contribution to $\chi$ from the power-law with
  exponential cutoff, c) Contribution to $\chi$ from the sphere+disk
  model.}
\label{fig:hextealone}
\end{figure*}

\begin{figure*}
\centerline{
\psfig{figure=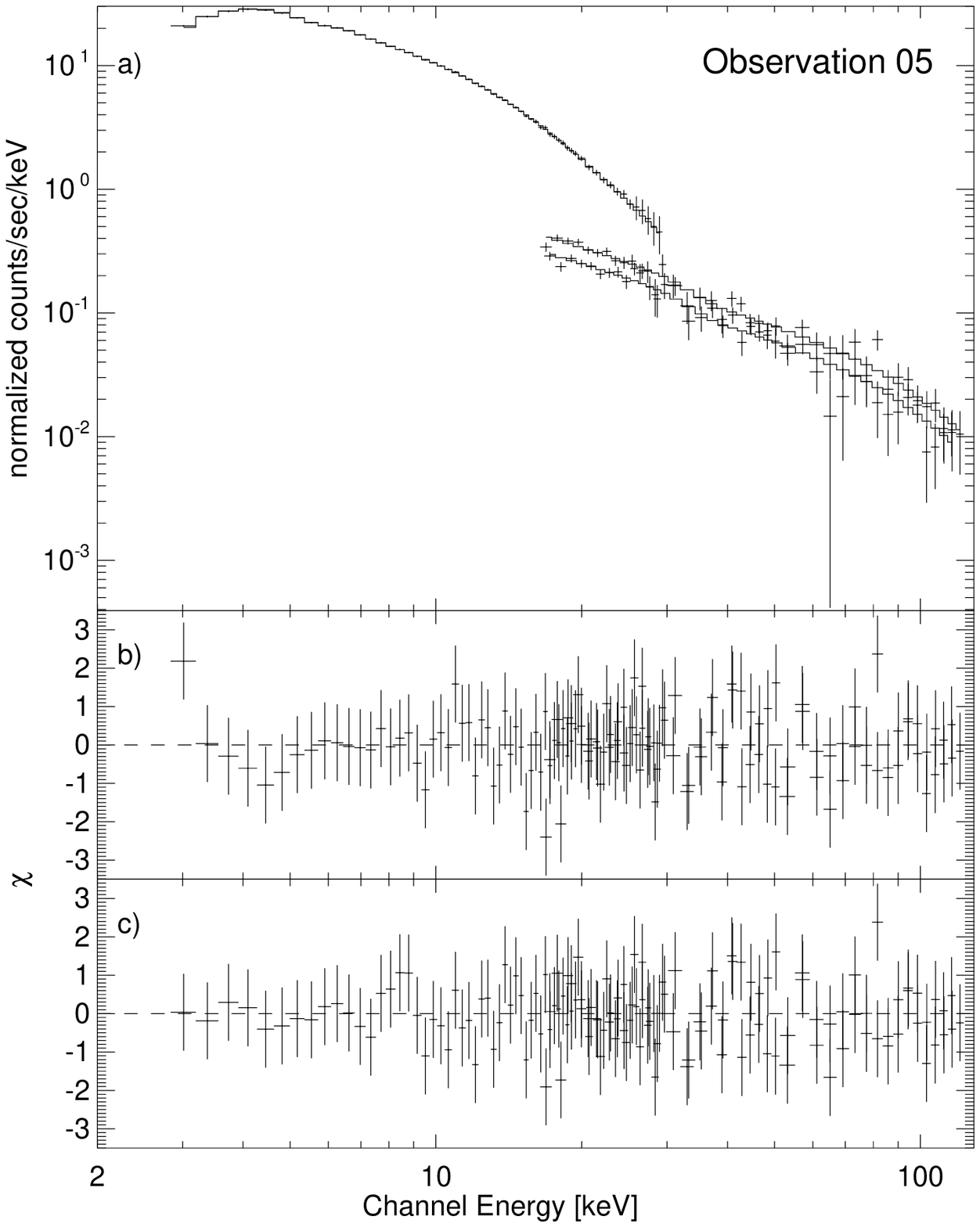,width=0.45\textwidth}
\psfig{figure=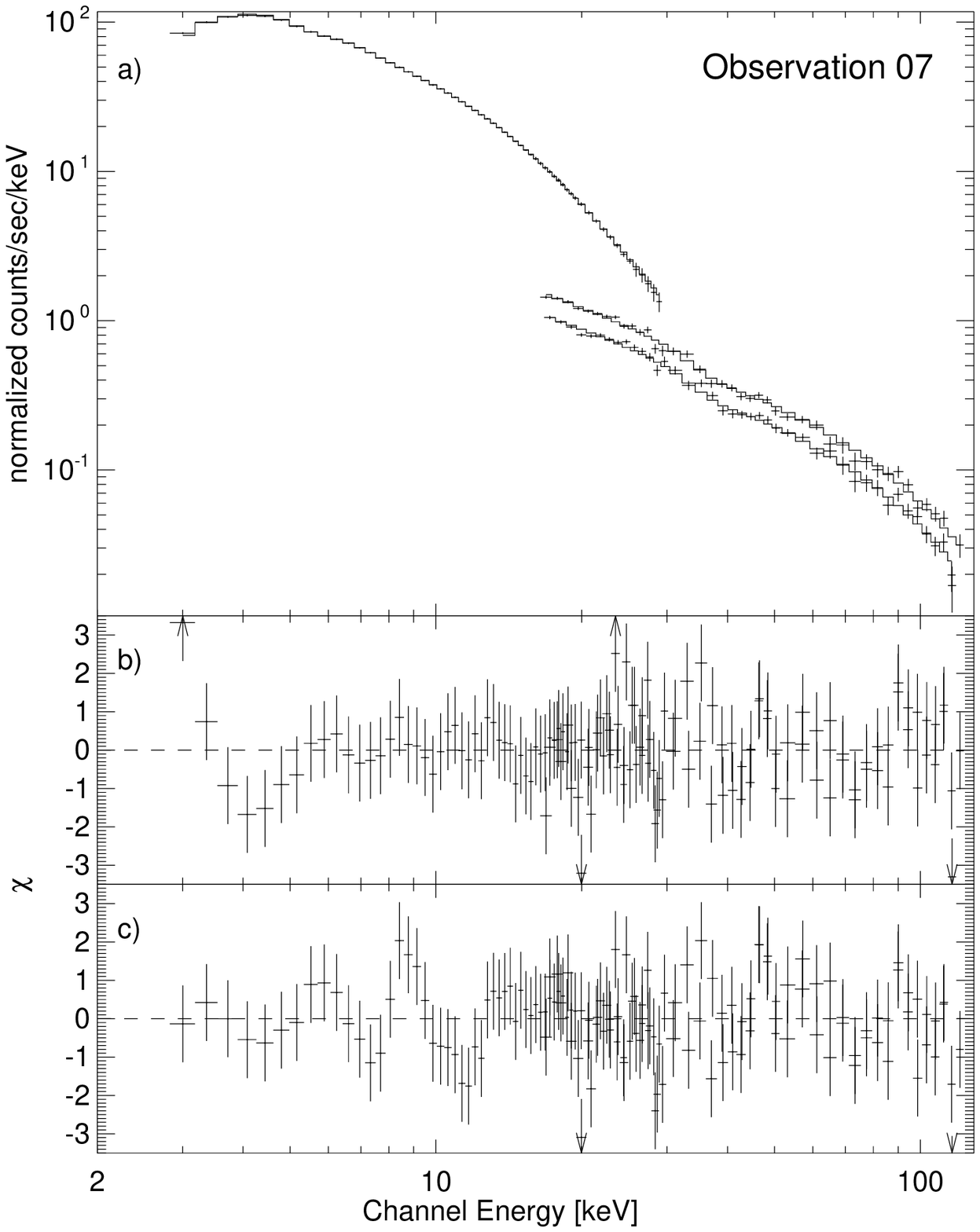,width=0.45\textwidth}}
\caption{\small Spectral modeling of the joint PCA and HEXTE data for
  Observation~5 and~7. Residues are shown as the contribution to $\chi$. a)
  Count rate spectrum and the best fit broken power-law with Gaussian line.
  b) Contribution to $\chi$ from the best fit broken power-law with a
  Gaussian line (parameters not given in text). c) Contribution to $\chi$
  from the best fit ionized reflector model with a Gaussian line.}
\label{fig:jointfit}
\end{figure*}

For our analysis of the RXTE broad band spectrum, we used several different
spectral models consistent with the range of parameterizations currently
used in the literature to describe the spectra of BHC.  As in the ASCA
analysis (\S\ref{sec:asca}), we fixed $N_{\rm H} = 6 \times 10^{21}~{\rm
  cm^2}$. We first used the purely phenomenological exponentially cutoff
power law and broken power law models as a broad characterization of the
data. The results of this modeling are given in \S\ref{sec:phenom} and in
Tables~\ref{tab:jwstud} and \ref{tab:jwstudb}.  We then applied the three
more physically motivated models that are currently discussed in the
literature: reflection of a power law off an (ionized) accretion disk
(\S\ref{sec:reflect} and Table~\ref{tab:jwstud2}), `sphere and disk' corona
Comptonization models (\S\ref{sec:corona} and Table~\ref{tab:jwstud3}), and
ADAF models (\S\ref{sec:adaf}). The ADAF models are only applied to the
unfolded data. Residuals for each of the best fits to the data from
Observations~5 and~7 are shown in
Figures~\ref{fig:pcaalone},~\ref{fig:hextealone}, and \ref{fig:jointfit}
(except for the ADAF models, where we present Observations 1 and 5).  We
chose to present these former two observations because not only are they at
extremes in terms of observed luminosity (Observation~5 is the faintest,
and Observation~7 is the second brightest, Table~\ref{tab:log}), but also
because they show detectable differences in their timing properties
(paper~II).

\subsubsection{Phenomenological Models}\label{sec:phenom}

Results from the purely phenomenological fits to the data, i.e., the broken
power law and exponentially cutoff power law, are presented in
Tables~\ref{tab:jwstud} and \ref{tab:jwstudb}. We see that a broken power
law plus a Gaussian models the PCA data very well.  The low
$\chi^2_{\rm red}$ (0.15--0.32) indicates that to some extent we may be
fitting systematic features in the PCA response.  The same may be true for
the best fit parameters of the Fe line.  The line widths ($\sigma \approx
0.6$\,keV) and equivalent widths ($\approx 130$\,eV) are larger than for
the ASCA observations, with the exception of Observation~5 which has line
parameters comparable to the ASCA observations.  As Observation~5 has the
lowest count rate, it is more dominated by statistical errors and less
dominated by systematic errors than the other observations. Even ignoring
the possible systematic effects, however, we see that any observed line is
narrower and weaker than is commonly observed in AGN.

\begin{table*}
\caption{\small Parameters for Gaussian line plus broken power law models
  (PCA only).}
{\small 
%
%\begin{sideways}
%\begin{minipage}{0.95\textheight}

\begin{center}
{Model: $\exp(-N_{\rm H}\sigma_a)\left( {\sf gauss}(E_{\rm
      line},\sigma,A_{\rm line}) + {\sf bknpower}(\Gamma_1,\Gamma_2,E_{\rm
      break},A_{\rm bkn}) \right)$}\\[3mm] 

\begin{tabular}{ccccccccccc}
\hline
\hline
\noalign{\vspace*{0.7mm}}
 Obs  & $E_{\rm line}$ & $\sigma $ &
      $A_{\rm line}$ & EW & $\Gamma_1    $ &
      $E_{\rm break}$ & $\Gamma_2$ & $A_{\rm bkn} $ 
      & $\chi^2/{\rm dof}$ & $\chi^2_{\rm red}$ \\ 
      & $\,({\rm keV})$ & $\,({\rm keV})$ & ($\times 10^{-3}$)  & $\,({\rm
      eV})$ & & $\,({\rm keV})$  &  & \\
\noalign{\vspace*{0.7mm}}
\hline
\noalign{\vspace*{0.7mm}}
01 & $\errtwa{    6.48}{    0.14}{    0.15}$ & $\errtwa
{     0.6}{     0.2}{     0.2}$ & $\errtwa{    1.97}{    0.42}{    0.36}$ 
& $\errtwa{ 130}{24}{22}$ &
$\errtwa{    1.80}{    0.01}{    0.01}$ & $\errtwa{    11.2}{     0.3}
{     0.4}$ & $\errtwa{    1.53}{    0.02}{    0.02}$ & $\errtwa{    0.44}
{    0.01}{    0.01}$ &    13.6/ 52 &    0.26 \\  
\noalign{\vspace*{0.7mm}}
\noalign{\vspace*{0.7mm}}
02 & $\errtwa{    6.47}{    0.15}{    0.16}$ & $\errtwa
{     0.6}{     0.2}{     0.2}$ & $\errtwa{    1.73}{   0.40}{    0.35}$ 
& $\errtwa{129}{ 26}{24}$ &
$\errtwa{    1.80}{    0.01}{    0.01}$ & $\errtwa{    10.9}{     0.4}
{     0.4}$ & $\errtwa{    1.53}{    0.02}{    0.02}$ & $\errtwa{    0.38}
{    0.01}{    0.01}$ &    16.6/52 &    0.32 \\  
\noalign{\vspace*{0.7mm}}
\noalign{\vspace*{0.7mm}}
03 & $\errtwa{    6.47}{    0.14}{    0.15}$ & $\errtwa
{     0.5}{     0.2}{     0.2}$ & $\errtwa{    1.55}{    0.29}{    0.31}$ 
& $\errtwa{121}{24}{23}$ &
$\errtwa{    1.79}{    0.01}{    0.01}$ & $\errtwa{    10.9}{     0.4}
{     0.4}$ & $\errtwa{    1.53}{    0.02}{    0.03}$ & $\errtwa{    0.37}
{    0.01}{    0.01}$ &   13.5/ 52 &    0.26 \\  

\noalign{\vspace*{0.7mm}}
\noalign{\vspace*{0.7mm}}
04 & $\errtwa{    6.45}{    0.08}{    0.15}$ & $\errtwa
{     0.5}{     0.2}{     0.2}$ & $\errtwa{    1.05}{    0.25}{    0.22}$ &
$\errtwa{121}{28}{23}$ &
$\errtwa{    1.78}{    0.01}{    0.01}$ & $\errtwa{    10.8}{     0.4}
{     0.5}$ & $\errtwa{    1.54}{    0.03}{    0.03}$ & $\errtwa{    0.24}
{    0.00}{    0.00}$ &   20.7/ 52 &    0.38 \\  
\noalign{\vspace*{0.7mm}}
\noalign{\vspace*{0.7mm}}
05 & $\errtwa{    6.43}{    0.15}{    0.17}$ & $\errtwa
{     0.2}{     0.3}{     0.2}$ & $\errtwa{    0.31}{    0.09}{    0.08}$ 
& $\errtwa{84}{25}{21}$ &
$\errtwa{    1.72}{    0.01}{    0.01}$ & $\errtwa{    10.8}{     0.8}
{     0.8}$ & $\errtwa{    1.49}{    0.04}{    0.05}$ & $\errtwa{    0.09}
{    0.00}{    0.00}$ &   22.9/ 52 &    0.44 \\  
\noalign{\vspace*{0.7mm}}
\noalign{\vspace*{0.7mm}}
06 & $\errtwa{    6.40}{    0.14}{    0.15}$ & $\errtwa
{     0.5}{     0.2}{     0.2}$ & $\errtwa{    1.32}{    0.28}{    0.27}$ 
& $\errtwa{123}{23}{23}$ &
$\errtwa{    1.80}{    0.01}{    0.01}$ & $\errtwa{    10.9}{     0.4}
{     0.4}$ & $\errtwa{    1.51}{    0.03}{    0.03}$ & $\errtwa{    0.30}
{    0.01}{    0.01}$ &    25.2/ 52 &    0.48 \\  
\noalign{\vspace*{0.7mm}}
\noalign{\vspace*{0.7mm}}
07 & $\errtwa{    6.45}{    0.13}{    0.13}$ & $\errtwa
{     0.6}{     0.2}{     0.2}$ & $\errtwa{    1.92}{    0.38}{    0.34}$ 
& $\errtwa{140}{23}{23}$ &
$\errtwa{    1.83}{    0.01}{    0.01}$ & $\errtwa{    10.9}{     0.3}
{     0.4}$ & $\errtwa{    1.55}{    0.02}{    0.02}$ & $\errtwa{    0.42}
{    0.01}{    0.01}$ &    27.3/ 52 &    0.53 \\  
\noalign{\vspace*{0.7mm}}
\noalign{\vspace*{0.7mm}}
08 & $\errtwa{    6.40}{    0.14}{    0.15}$ & $\errtwa
{     0.5}{     0.2}{     0.2}$ & $\errtwa{    1.20}{    0.27}{    0.23}$ 
& $\errtwa{130}{26}{23}$ &
$\errtwa{    1.79}{    0.01}{    0.01}$ & $\errtwa{    10.8}{     0.5}
{     0.5}$ & $\errtwa{    1.54}{    0.03}{    0.03}$ & $\errtwa{    0.26}
{    0.01}{    0.01}$ &   28.3/ 52 &    0.54 \\ 
\noalign{\vspace*{0.7mm}}
\hline
\end{tabular}
\end{center}
%\end{minipage}
%\end{sideways}

%%% Local Variables: 
%%% mode: latex
%%% TeX-master: "
"
%%% End: 

}
\label{tab:jwstud}
\end{table*}

\begin{table*}
\caption{\small Parameters for
  exponentially cutoff power law models (HEXTE
  only).}

{\small 
\begin{center}
{Model: ${\sf const.}\left( A_{\rm PL} E^{-\Gamma} \exp\left(-E/E_{\rm cut}\right) \right)$}\\[3mm]

\begin{tabular}{cccccccc}
\hline
\hline
\noalign{\vspace*{0.7mm}}
 Obs  & ${\rm const.}$ & $\Gamma      $ & $E_{\rm cut} $ & $A_{\rm PL}  $ &
      ${\rm const.}$& $\chi^2/{\rm dof}$ & $\chi^2_{\rm red}$ \\ 
      & & & $\,({\rm keV})$ & & & &                    \\ 
\noalign{\vspace*{0.7mm}}
\hline
\noalign{\vspace*{0.7mm}}
01 & ${\it     1.00}$ & $\errtwa{ 1.25}{ 0.06}{ 0.06}$ & $\errtwa{    101}
{     18}{     14}$ & $\errtwa{   0.082}{   0.014}{   0.012}$ & $\errtwa
{    0.99}{    0.01}{    0.01}$ &   69.2/ 80 &    0.87 \\  
\noalign{\vspace*{0.7mm}}
02 & ${\it     1.00}$ & $\errtwa{ 1.12}{ 0.08}{ 0.08}$ & $\errtwa{     79}
{     13}{     10}$ & $\errtwa{   0.052}{   0.011}{   0.009}$ & $\errtwa
{    0.92}{    0.02}{    0.02}$ &   69.0/ 80 &    0.86 \\  
\noalign{\vspace*{0.7mm}}
03 & ${\it     1.00}$ & $\errtwa{ 1.16}{ 0.17}{ 0.18}$ & $\errtwa{     94}
{     63}{     28}$ & $\errtwa{   0.052}{   0.029}{   0.019}$ & $\errtwa
{    1.08}{    0.04}{    0.04}$ &   71.8/ 80 &    0.90 \\  
\noalign{\vspace*{0.7mm}}
04 & ${\it     1.00}$ & $\errtwa{ 1.15}{ 0.11}{ 0.11}$ & $\errtwa{     85}
{     25}{     16}$ & $\errtwa{   0.035}{   0.011}{   0.009}$ & $\errtwa
{    0.98}{    0.02}{    0.02}$ &   76.2/ 80 &    0.95 \\  
\noalign{\vspace*{0.7mm}}
05 & ${\it     1.00}$ & $\errtwa{ 1.18}{ 0.19}{ 0.23}$ & $\errtwa{    115}
{     85}{     46}$ & $\errtwa{   0.016}{   0.011}{   0.007}$ & $\errtwa
{    0.99}{    0.05}{    0.05}$ &   71.3/ 80 &    0.89 \\  
\noalign{\vspace*{0.7mm}}
06 & ${\it     1.00}$ & $\errtwa{ 1.19}{ 0.09}{ 0.09}$ & $\errtwa{    103}
{     29}{     19}$ & $\errtwa{   0.049}{   0.012}{   0.010}$ & $\errtwa
{    0.93}{    0.02}{    0.02}$ &   87.9/ 80 &    1.10 \\  
\noalign{\vspace*{0.7mm}}
07 & ${\it     1.00}$ & $\errtwa{ 1.21}{ 0.07}{ 0.08}$ & $\errtwa{     95}
{     20}{     14}$ & $\errtwa{   0.066}{   0.014}{   0.012}$ & $\errtwa
{    0.96}{    0.02}{    0.02}$ &  101.5/ 80 &    1.27 \\  
\noalign{\vspace*{0.7mm}}
08 & ${\it     1.00}$ & $\errtwa{ 1.08}{ 0.10}{ 0.10}$ & $\errtwa{     81}
{     19}{     13}$ & $\errtwa{   0.031}{   0.009}{   0.007}$ & $\errtwa
{    0.98}{    0.02}{    0.02}$ &  106.6/ 80 &    1.33 \\  
\noalign{\vspace*{0.7mm}}
\hline
\end{tabular}
\end{center}
}

%%% Local Variables: 
%%% mode: latex
%%% TeX-master: t
%%% End: 

\label{tab:jwstudb}
\end{table*}

The $\approx 3$--$10$\,keV spectral power-law slope is close to the
`canonical value' of $\Gamma = 1.7$; however, the PCA shows evidence for a
hardening of this spectral slope above $\approx 10$\,keV.  HEXTE data alone
also show the high energy spectrum to be harder than the 3--10\,keV
spectrum (Fig.~\ref{fig:hextealone} and Table~\ref{tab:jwstudb}).  Note
that the difference between the PCA and HEXTE photon indices is
\emph{greater} than the discrepancy between the PCA and HEXTE best fit Crab
photon indices (appendix~\ref{sec:rxteapp}), and therefore it is unlikely
to be a systematic effect.

Such a hard HEXTE spectrum is consistent with previous observations by the
Oriented Scintillation Spectrometer Experiment (OSSE) on board the Compton
Gamma-ray Observatory (CGRO) (\cite{grabelsky:95a}; see also
\cite{zdziarski:98a}).  \citey{grabelsky:95a} found a slightly harder
photon index of $\Gamma = 0.88$ and an exponential cutoff of $E_{\rm cut} =
68$\,keV, somewhat lower than observed here.  Note, however, that the OSSE
observations extended to $\approx 500$\,keV as opposed to the $\sim
110$\,keV of our HEXTE observation. Therefore, the HEXTE data for
GX~339$-$4 do not strongly constrain the exponential rollover, and slightly
harder power laws with lower exponential cutoffs are permitted.

\subsubsection{Reflection Models}\label{sec:reflect}

A spectral hardening above $\approx 7$\,keV is the expected signature of
reflection of a hard continuum off of cold material (\cite{magdziarz:95a}).
\citey{ueda:94a} applied reflection models to \textsl{Ginga} data of
GX~339$-$4 and found strong evidence of reflection, whereas
\citey{grabelsky:95a} found no evidence of reflection in OSSE data.
\citey{zdziarski:98a} jointly fit these simultaneously observed data sets
and find that reflection models, albeit with a large Fe abundance, provide
a very good description of the data. We have applied the models of
\citey{magdziarz:95a}, as implemented in XSPEC (\emph{pexrav},
\emph{pexriv}), to the GX~339$-$4 data. These models consider an
exponentially cutoff power law reflected off of neutral (\emph{pexrav}) or
partially ionized (\emph{pexriv}) cold material.

\begin{table*}
\caption{\small Parameters for Gaussian line plus multicolor disk plus
  ionized reflection models after Magdziarz \& Zdziarski
  (1995). \label{tab:jwstud2}}   
 {\small
 \begin{sideways} \begin{minipage}{0.95\textheight}
 \begin{center}
{Model: ${\sf const.} \cdot      \exp(-N_{\rm H}\sigma_{a}) \left
    (     {\sf diskbb}(A_{\rm dbb},kT_{\rm dbb}) +      {\sf pexriv}(E_{\rm
      fold},\Gamma,f,z,A_{\rm X},     A_{\rm Fe},\cos i,     A_{\rm pex}) +
    {\sf gauss}(E_{\rm line},\sigma,     A_{\rm line}) \right ) $} \nl 
 \begin{tabular}{ccccccccccccccc}
\hline
\hline
\noalign{\vspace*{0.7mm}}
Obs & $A_{\rm dbb}$ & $\Gamma$ & $\Gamma_{\rm HEXTE}$      & $f$ & $A_{\rm
  Fe}$ & $\xi$      &  $A_{\rm pex}$ &     $E_{\rm line}$ & $A_{\rm line}$
& EW & const. & const.      & $\chi^2/$dof & $\chi^2_{\rm red}$ \nl      &
$(\times 10^{5})$ & & & & & $({\rm erg~cm~s^{-1}})$ & & (keV)      &
  ($\times 10^{-4}$) & (eV) \nl   
\noalign{\vspace*{0.7mm}}
\hline
\noalign{\vspace*{0.7mm}}
01& $\errtwa{ 1.64}{ 0.82}{ 0.83}$     & $\errtwa{ 1.81}{ 0.02}{ 0.02}$ & --- & $ \errtwa{ 0.41}{ 0.06}{ 0.05}$ & $ \errtwa{ 1.54}{ 0.67}{ 0.40}$ & $ \errtwa{  78.2}{  28.7}{  23.7}$ & $ \errtwa{ 0.43}{ 0.01}{ 0.01}$ & $ \errtwa{ 6.22}{ 0.22}{ 0.23}$ & $ \errtwa{ 7.00}{ 2.95}{ 3.10}$ & $ \errtwa{  41}{  18}{  18}$ & --- & --- & $  25./51$ &  0.49 \nl
\noalign{\vspace*{0.7mm}}
01& $ \errtwa{ 1.54}{ 0.81}{ 0.82}$      & $\errtwa{ 1.81}{ 0.02}{ 0.02}$ & $\errtwa{ 1.76}{ 0.04}{ 0.03}$ & $ \errtwa{ 0.47}{ 0.07}{ 0.06}$ & $ \errtwa{ 2.26}{ 1.01}{ 0.65}$ & $ \errtwa{  60.5}{  20.8}{  21.8}$ & $ \errtwa{ 5.85}{ 2.75}{ 2.98}$ &  {\it 6.4} & $ \errtwa{ 5.85}{ 2.75}{ 2.98}$ & $ \errtwa{  35}{  18}{  17}$ & $ \errtwa{ 0.57}{ 0.09}{ 0.07}$ & $ \errtwa{ 0.57}{ 0.09}{ 0.07}$ & $ 113./132$ &  0.86 \nl
\noalign{\vspace*{0.7mm}}
\noalign{\vspace*{0.7mm}}
02& $\errtwa{ 1.68}{ 0.73}{ 0.73}$     & $\errtwa{ 1.81}{ 0.02}{ 0.02}$ & --- & $ \errtwa{ 0.41}{ 0.06}{ 0.05}$ & $ \errtwa{ 1.36}{ 0.58}{ 0.36}$ & $ \errtwa{  82.2}{  31.6}{  23.9}$ & $ \errtwa{ 0.37}{ 0.01}{ 0.01}$ & $ \errtwa{ 6.16}{ 0.23}{ 0.25}$ & $ \errtwa{ 5.96}{ 2.60}{ 2.82}$ & $ \errtwa{  40}{  17}{  19}$ & --- & --- & $  21./51$ &  0.41 \nl
\noalign{\vspace*{0.7mm}}
02& $ \errtwa{ 1.49}{ 0.73}{ 0.73}$      & $\errtwa{ 1.80}{ 0.02}{ 0.02}$ & $\errtwa{ 1.75}{ 0.05}{ 0.04}$ & $ \errtwa{ 0.52}{ 0.09}{ 0.07}$ & $ \errtwa{ 2.46}{ 1.50}{ 0.80}$ & $ \errtwa{  54.0}{  21.3}{  22.5}$ & $ \errtwa{ 4.08}{ 2.45}{ 2.68}$ &  {\it 6.4} & $ \errtwa{ 4.08}{ 2.45}{ 2.68}$ & $ \errtwa{  18}{  28}{  10}$ & $ \errtwa{ 0.57}{ 0.12}{ 0.08}$ & $ \errtwa{ 0.52}{ 0.11}{ 0.07}$ & $ 124./132$ &  0.94 \nl
\noalign{\vspace*{0.7mm}}
\noalign{\vspace*{0.7mm}}
03& $\errtwa{ 1.34}{ 0.70}{ 0.69}$     & $\errtwa{ 1.81}{ 0.02}{ 0.02}$ & --- & $ \errtwa{ 0.42}{ 0.06}{ 0.05}$ & $ \errtwa{ 1.49}{ 0.62}{ 0.42}$ & $ \errtwa{  59.8}{  24.1}{  24.8}$ & $ \errtwa{ 0.36}{ 0.01}{ 0.01}$ & $ \errtwa{ 6.22}{ 0.24}{ 0.27}$ & $ \errtwa{ 5.56}{ 2.53}{ 2.66}$ & $ \errtwa{  39}{  16}{  18}$ & --- & --- & $  14./51$ &  0.28 \nl
\noalign{\vspace*{0.7mm}}
03& $ \errtwa{ 1.23}{ 0.70}{ 0.70}$      & $\errtwa{ 1.81}{ 0.02}{ 0.02}$ & $\errtwa{ 1.68}{ 0.05}{ 0.05}$ & $ \errtwa{ 0.44}{ 0.07}{ 0.06}$ & $ \errtwa{ 1.61}{ 0.76}{ 0.46}$ & $ \errtwa{  50.9}{  22.6}{  22.9}$ & $ \errtwa{ 4.96}{ 2.46}{ 2.56}$ &  {\it 6.4} & $ \errtwa{ 4.96}{ 2.46}{ 2.56}$ & $ \errtwa{  36}{  19}{  18}$ & $ \errtwa{ 0.42}{ 0.09}{ 0.07}$ & $ \errtwa{ 0.45}{ 0.09}{ 0.07}$ & $  89./132$ &  0.68 \nl
\noalign{\vspace*{0.7mm}}
\noalign{\vspace*{0.7mm}}
04& $\errtwa{ 1.13}{ 0.47}{ 0.48}$     & $\errtwa{ 1.79}{ 0.02}{ 0.02}$ & --- & $ \errtwa{ 0.37}{ 0.07}{ 0.06}$ & $ \errtwa{ 1.33}{ 0.77}{ 0.43}$ & $ \errtwa{  70.2}{  31.3}{  27.3}$ & $ \errtwa{ 0.23}{ 0.00}{ 0.01}$ & $ \errtwa{ 6.17}{ 0.23}{ 0.23}$ & $ \errtwa{ 4.56}{ 1.80}{ 1.90}$ & $ \errtwa{  47}{  18}{  19}$ & --- & --- & $  15./51$ &  0.30 \nl
\noalign{\vspace*{0.7mm}}
04& $ \errtwa{ 1.03}{ 0.47}{ 0.47}$      & $\errtwa{ 1.79}{ 0.02}{ 0.02}$ & $\errtwa{ 1.73}{ 0.07}{ 0.05}$ & $ \errtwa{ 0.46}{ 0.12}{ 0.08}$ & $ \errtwa{ 2.26}{ 1.78}{ 0.86}$ & $ \errtwa{  44.2}{  24.1}{  23.7}$ & $ \errtwa{ 3.61}{ 1.60}{ 1.88}$ &  {\it 6.4} & $ \errtwa{ 3.61}{ 1.60}{ 1.88}$ & $ \errtwa{  39}{  18}{  20}$ & $ \errtwa{ 0.54}{ 0.14}{ 0.09}$ & $ \errtwa{ 0.52}{ 0.14}{ 0.08}$ & $ 100./132$ &  0.76 \nl
\noalign{\vspace*{0.7mm}}
\noalign{\vspace*{0.7mm}}
05& $\errtwa{ 0.39}{ 0.23}{ 0.24}$     & $\errtwa{ 1.73}{ 0.03}{ 0.02}$ & --- & $ \errtwa{ 0.43}{ 0.49}{ 0.13}$ & $ \errtwa{ 2.43}{11.04}{ 1.35}$ & $ \errtwa{  10.1}{  35.8}{  10.1}$ & $ \errtwa{ 0.09}{ 0.00}{ 0.00}$ & $ \errtwa{ 6.32}{ 0.27}{ 0.27}$ & $ \errtwa{ 1.84}{ 1.06}{ 0.94}$ & $ \errtwa{  48}{  29}{  25}$ & --- & --- & $  18./51$ &  0.34 \nl
\noalign{\vspace*{0.7mm}}
05& $ \errtwa{ 0.36}{ 0.22}{ 0.23}$      & $\errtwa{ 1.73}{ 0.03}{ 0.02}$ & $\errtwa{ 1.65}{ 0.12}{ 0.08}$ & $ \errtwa{ 0.44}{ 0.24}{ 0.12}$ & $ \errtwa{ 2.41}{ 3.49}{ 1.22}$ & $ \errtwa{   6.1}{  25.2}{   6.1}$ & $ \errtwa{ 1.87}{ 1.09}{ 0.96}$ &  {\it 6.4} & $ \errtwa{ 1.87}{ 1.09}{ 0.96}$ & $ \errtwa{  50}{  30}{  48}$ & $ \errtwa{ 0.49}{ 0.24}{ 0.12}$ & $ \errtwa{ 0.49}{ 0.23}{ 0.12}$ & $  88./132$ &  0.67 \nl
\noalign{\vspace*{0.7mm}}
\noalign{\vspace*{0.7mm}}
06& $\errtwa{ 1.61}{ 0.57}{ 0.57}$     & $\errtwa{ 1.80}{ 0.02}{ 0.02}$ & --- & $ \errtwa{ 0.47}{ 0.10}{ 0.07}$ & $ \errtwa{ 1.87}{ 1.05}{ 0.56}$ & $ \errtwa{  55.0}{  24.0}{  24.2}$ & $ \errtwa{ 0.29}{ 0.00}{ 0.01}$ & $ \errtwa{ 6.08}{ 0.22}{ 0.23}$ & $ \errtwa{ 5.45}{ 2.16}{ 2.30}$ & $ \errtwa{  45}{  18}{  19}$ & --- & --- & $  17./51$ &  0.33 \nl
\noalign{\vspace*{0.7mm}}
06& $ \errtwa{ 1.46}{ 0.55}{ 0.55}$      & $\errtwa{ 1.81}{ 0.02}{ 0.02}$ & $\errtwa{ 1.71}{ 0.05}{ 0.04}$ & $ \errtwa{ 0.52}{ 0.10}{ 0.07}$ & $ \errtwa{ 2.38}{ 1.26}{ 0.74}$ & $ \errtwa{  36.7}{  22.0}{  19.2}$ & $ \errtwa{ 4.06}{ 2.08}{ 2.05}$ &  {\it 6.4} & $ \errtwa{ 4.06}{ 2.08}{ 2.05}$ & $ \errtwa{  26}{  30}{   8}$ & $ \errtwa{ 0.50}{ 0.10}{ 0.07}$ & $ \errtwa{ 0.47}{ 0.09}{ 0.07}$ & $ 113./132$ &  0.86 \nl
\noalign{\vspace*{0.7mm}}
\noalign{\vspace*{0.7mm}}
07& $\errtwa{ 2.29}{ 0.73}{ 0.73}$     & $\errtwa{ 1.83}{ 0.02}{ 0.02}$ & --- & $ \errtwa{ 0.42}{ 0.06}{ 0.05}$ & $ \errtwa{ 1.37}{ 0.58}{ 0.37}$ & $ \errtwa{  84.8}{  32.5}{  23.9}$ & $ \errtwa{ 0.40}{ 0.01}{ 0.01}$ & $ \errtwa{ 6.16}{ 0.18}{ 0.19}$ & $ \errtwa{ 8.36}{ 2.67}{ 2.96}$ & $ \errtwa{  54}{  18}{  19}$ & --- & --- & $  25./51$ &  0.48 \nl
\noalign{\vspace*{0.7mm}}
07& $ \errtwa{ 2.10}{ 0.72}{ 0.72}$      & $\errtwa{ 1.83}{ 0.02}{ 0.02}$ & $\errtwa{ 1.74}{ 0.04}{ 0.03}$ & $ \errtwa{ 0.48}{ 0.07}{ 0.06}$ & $ \errtwa{ 1.97}{ 0.93}{ 0.57}$ & $ \errtwa{  61.5}{  21.9}{  22.7}$ & $ \errtwa{ 6.23}{ 2.76}{ 2.48}$ &  {\it 6.4} & $ \errtwa{ 6.23}{ 2.76}{ 2.48}$ & $ \errtwa{  43}{  20}{  17}$ & $ \errtwa{ 0.50}{ 0.08}{ 0.06}$ & $ \errtwa{ 0.48}{ 0.08}{ 0.06}$ & $ 144./132$ &  1.09 \nl
\noalign{\vspace*{0.7mm}}
\noalign{\vspace*{0.7mm}}
08& $\errtwa{ 1.25}{ 0.47}{ 0.47}$     & $\errtwa{ 1.80}{ 0.02}{ 0.02}$ & --- & $ \errtwa{ 0.38}{ 0.06}{ 0.05}$ & $ \errtwa{ 1.36}{ 0.67}{ 0.42}$ & $ \errtwa{  72.6}{  30.7}{  26.6}$ & $ \errtwa{ 0.25}{ 0.01}{ 0.01}$ & $ \errtwa{ 6.15}{ 0.19}{ 0.20}$ & $ \errtwa{ 5.55}{ 1.95}{ 1.87}$ & $ \errtwa{  55}{  19}{  18}$ & --- & --- & $  18./51$ &  0.35 \nl
\noalign{\vspace*{0.7mm}}
08& $ \errtwa{ 1.14}{ 0.47}{ 0.47}$      & $\errtwa{ 1.80}{ 0.02}{ 0.02}$ & $\errtwa{ 1.71}{ 0.07}{ 0.05}$ & $ \errtwa{ 0.50}{ 0.15}{ 0.09}$ & $ \errtwa{ 2.81}{ 2.23}{ 1.11}$ & $ \errtwa{  39.0}{  24.8}{  21.7}$ & $ \errtwa{ 4.14}{ 1.89}{ 1.72}$ &  {\it 6.4} & $ \errtwa{ 4.14}{ 1.89}{ 1.72}$ & $ \errtwa{  44}{  20}{  19}$ & $ \errtwa{ 0.51}{ 0.14}{ 0.09}$ & $ \errtwa{ 0.50}{ 0.14}{ 0.09}$ & $ 141./132$ &  1.07 \nl
\noalign{\vspace*{0.7mm}}
\noalign{\vspace*{0.7mm}}
\hline
 \end{tabular}
 \end{center} \end{minipage} \end{sideways} } 

%%% Local Variables: 
%%% mode: latex
%%% TeX-master: t
%%% End: 

\vskip -0.5 true in
\tablecomments{Models have been fit to: PCA data
  only (51~dof), and PCA and HEXTE data where the photon index of the
  reflected power law has been allowed to differ between the PCA and the
  HEXTE data (132~dof). Fit parameters are: the normalization of the
  multicolor disk blackbody, $A_{\rm dbb}$ (the disk temperature has been
  fixed to 250\,eV); the power law slope, $\Gamma$;
  cutoff energy, $E_{\rm fold}$; relative reflection fraction, $f\equiv
  \Delta \Omega/2 \pi$; Fe abundance relative to solar, $A_{\rm Fe}$; 
  the disk ionization parameter, $\xi\equiv$ luminosity/(density $\times$
  radius$^2$); and a normalization, $A_{\rm pex}$.  The ionized 
  reflection models also have parameters for the temperature of the disk,
  $T_{\rm disk}$ (fixed at $10^6$\,K); abundances of elements heavier than
  He relative to solar, $A_{\rm X}$ (fixed at unity); and disk inclination
  angle, $i$ (fixed at $45^\circ$). Parameters typeset in italics 
  have been held constant for that particular fit. The iron line width was
  fixed at 0.1\,keV. }
\end{table*}

In Table~\ref{tab:jwstud2} we show the fit results for reflection off of
partially ionized material similar to the models presented by
\citey{zdziarski:98a}.  Just as in \citey{zdziarski:98a}, we include a disk
component where we fix the inner disk temperature to 250\,eV.  As PCA does
not usefully constrain models below 3\,keV, the disk component is not
strongly constrained; typically the $\chi^2$ values were higher by 5--20
without this component.  We also fix the reflector inclination angle at
$45^\circ$, fix the disk temperature at $T_{\rm disk} = 10^6$\,K, freeze
the abundances at solar, but let the Fe abundance be a free parameter.  In
all our fits we found that the Gaussian line width, $\sigma$, would tend to
drift toward $0$, so we fixed $\sigma = 0.1$\,keV.  For the combined PCA
and HEXTE data, we also fixed the Gaussian line energy to 6.4\,keV.  For
fits to PCA data alone and joint PCA/HEXTE data, the exponential cutoff
energy, $E_{\rm fold}$, would drift towards very large energy ($\gg
1000$\,keV). We therefore only considered pure power laws without cutoffs.
\citey{zdziarski:98a} have argued that the high energy cutoff is sharper
than exponential, which one would not expect to be strongly constrained by
the combined PCA/HEXTE data.

As for the \textsl{Ginga} data of GX~339$-$4 (\cite{ueda:94a}), the PCA
data alone were extremely well described by reflection models.  Again,
however, the extremely low $\chi^2_{\rm red}$ (as low as 0.28) makes us
caution that these models might partly be fitting systematics in the PCA
response.  PCA and \textsl{Ginga} are also very similar instruments in
terms of design, and so to some extent they should exhibit similar
systematic effects (as discussed in appendix~\ref{sec:rxteapp}, the
internal consistency of the PCA calibration is now as good as or better
than that for the \textsl{Ginga} calibration.).  Note that the best fit Fe
line equivalent widths here are significantly smaller than those found with
the purely phenomenological models discussed in \S\ref{sec:phenom}.

The fits for the HEXTE data alone (not shown) were similar to the OSSE
results of \citey{grabelsky:95a}.  Namely, if one allows for an exponential
cutoff (typically $\approx 100$\,keV) to the power law, the best fit
reflection fraction becomes $f \aproxlt 0.01$.  Such a small reflection
fraction is not surprising considering how well a pure exponentially cutoff
power law fits the HEXTE data (Table~\ref{tab:jwstudb}).  If one does
not allow an exponential cutoff, the reflection fraction becomes $f
\aproxgt 3$.  Such a fit is trying to mimic a hard power law with a
high energy cutoff.

A joint analysis of the PCA and the HEXTE data should be similar to a joint
analysis of the Ginga and OSSE data. Indeed, such an analysis yields
results comparable to those presented by \citey{zdziarski:98a} if we
constrain the photon index of the incident power law to be the same for
both the PCA and HEXTE data.  Notable for the results of such fits (not
presented) is the fairly large overabundance of Fe ($A_{\rm Fe} =
3.2$--$5.2$).  Similarly, \citey{zdziarski:98a} find a large $A_{\rm
  Fe}=2.5$--$3.0$ \emph{except} for a short data set, more likely dominated
by statistical errors rather than systematic errors, where they find
$A_{\rm Fe}=1.6$--$2.0$.  For our joint PCA/HEXTE data, the best-fit
reflection fractions were approximately 20\% larger than the best-fit
reflection fraction for PCA data alone.  Such an increase in reflection
fraction in general will reproduce the spectral hardening seen in the HEXTE
energy bands.  Increasing the average best fit Fe abundance from $\langle
A_{\rm Fe} \rangle = 1.6$ (PCA data alone) to $\langle A_{\rm Fe} \rangle =
4.0$ (joint PCA/HEXTE data) also leads to an increased spectral hardening
above $\approx 7$\,keV, while leaving the spectrum below $\approx 7$\,keV
relatively unchanged. That is, such a fit helps to reproduce the spectral
break at $\approx 10$\,keV.

For the joint PCA/HEXTE analysis, there is clearly a worry that these
results are influenced by the systematic differences between the PCA and the
HEXTE responses.  We therefore performed reflection model fits where we
constrained all fit parameters to be the same for the PCA and the HEXTE
data \emph{except} for the incident power law photon index, which we
allowed to vary between the two instruments\footnote{The photon index was
  constrained to be the same for HEXTE Cluster~A and~B. The necessary
  different normalizations between the PCA and the HEXTE models were
  subsumed into the constants multiplying the HEXTE models. As HEXTE
  requires a harder power law, these constants were now $0.42$--$0.57$, as
  opposed to $\approx 0.7$.  Furthermore, the constants showed larger
  uncertainties, as the uncertainty of the HEXTE photon index now couples
  strongly to the value of the constants.}.  Such models provided
reasonably good fits to the data, with $\chi^2_{\rm red}$ ranging from 0.67
to 1.09.  The difference between the PCA and the HEXTE best fit photon
indices ranged from 0.05 to 0.13, with an average value of 0.08.  This is
consistent with the systematic difference between the best-fit photon
indeces for the Crab spectrum. For these models we find $\langle A_{\rm Fe}
\rangle = 2.3$, which is more consistent with the results for fits to the
PCA data only, and is slightly smaller than the results found by
\citey{zdziarski:98a}.  Note that we also find smaller values of the
ionization parameter, $\xi$, than were found by \citey{zdziarski:98a}.

\subsubsection{Corona Models}\label{sec:corona}

We considered `sphere+disk' Comptonization models (\cite{dove:97b}) of the
GX~339$-$4 observations. We have previously applied these models to an RXTE
observation of Cygnus~X--1 (\cite{dove:98a}).  The models consist of a
central, spherical corona surrounded by a geometrically thin, flat disk.
Seed photons for Comptonization come from the disk, which has a radial
temperature distribution $kT_{\rm disk}(R) \propto R^{-3/4}$ and a
temperature of 150\,eV at the inner edge of the disk.  Hard flux from the
corona further leads to reflection features from the disk or to soft
photons due to thermalization of the hard radiation. The (non-uniform)
temperature and pair balance within the corona are self-consistently
calculated from the radiation field (\cite{dove:97b}).

As described by \citey{dove:97b}, we parameterize our models by the coronal
compactness
\begin{equation}
\ell_{\rm c} \equiv \frac{\sigma_{\rm T}}{m_{\rm e} c^3}\frac{L_{\rm
    C}}{R_{\rm C}} ~~, 
\label{eq:compact}
\end{equation}
where $\sigma_{\rm T}$ is the Thomson cross section, $m_{\rm e}$ is the
electron mass, $L_{\rm C}$ is the luminosity of the corona, and $R_{\rm C}$
is the radius of the corona. Likewise, we define a disk compactness,
$\ell_{\rm d} \equiv (1-f_{\rm c}) (\sigma_{\rm T}/m_{\rm e} c^3) P_{\rm
  G}/R_{\rm C}$, where $P_{\rm G}$ is the \emph{total} rate of
gravitational energy dissipated in the system, and $f_{\rm c}$ is the
fraction dissipated in the corona.  In calculating the numerical models, we
set $\ell_{\rm d} = 1$. Models with other values of $\ell_{\rm d}$ yield
the same ranges of self-consistent coronal temperatures and opacities.  In
general $f_{\rm c} = \ell_{\rm c} / (\ell_{\rm d} + \ell_{\rm c})$
(\cite{dove:97b}).  Based upon the `sphere+disk' geometry, a fraction $f
\approx 0.32$ of the coronal flux is absorbed by the disk
(\cite{dove:97b}).  The models are further parameterized by an initial
electron coronal optical depth, $\tau_{\rm c}$ (approximately equal to the
total optical depth, as pair production is negligible for the parameters of
interest to us), and a normalization constant $A_{\rm kot}$.  From the
best-fit compactness and optical depth, the average coronal temperature can
be calculated \emph{a posteriori}.

\begin{table*}
\caption{\small Models of `sphere+disk' Comptonization plus a Gaussian line
  fit to PCA data only and HEXTE data only. \label{tab:jwstud3}}  
\smallskip
{\small 
%
% PCA: kotelp with a Gaussian Line
%
\begin{tabular}{cccccccccccc}
\hline
\hline
\noalign{\vspace*{0.7mm}}
 Obs  & $E_{\rm Line}$ & $\sigma$ & $A_{\rm Line}$ & EW &
 $l_{\rm c}$ & $\tau_{\rm c}$ & $A_{\rm kot}$ & ${\rm const.}$
 & $kT_{\rm c}$ & $\chi^2/{\rm dof}$ &  $\chi^2_{\rm red}$ \\ 
      & (keV) & (keV) & ($\times 10^{-3}$) & (eV) & & & & & (keV) \\ 
\noalign{\vspace*{0.7mm}}
\hline
\noalign{\vspace*{0.7mm}}
01 & $\errtwa{    6.39}{    0.18}{    0.17}$ & $\errtwa
 {     0.8}{     0.2}{     0.1}$ & $\errtwa{    3.09}{    0.60}{    0.53}$
 & $\errtwa{197}{30}{30}$ & $\errtwa{     0.62}{     0.05}{     0.04}$ &
 $\errtwa{     3.3} 
 {     0.1}{     0.1}$ & $\errtwa{     2.38}{     0.06}{     0.08}$
 & \nodata & $\errtwa{28.6}{0.4}{0.4}$ &  19.2/53 &    0.36 \\  
\noalign{\vspace*{0.7mm}}
01 & \nodata & \nodata & \nodata & \nodata & $\errtwa{1.83}{0.17}{0.13}$ & $\errtwa{2.9}{0.5}{0.3}$ &
 $\errtwa{0.91}{0.09}{0.09}$ & $\errtwa{0.99}{0.01}{0.01}$ &
 $\errtwa{43.9}{8}{8}$ &  71.2/80 &
 0.89 \\  
\noalign{\vspace*{0.7mm}}
02 & $\errtwa{    6.36}{    0.19}{    0.19}$ & $\errtwa
 {     0.8}{     0.2}{     0.1}$ & $\errtwa{    2.69}{    0.57}{    0.50}$
 & $\errtwa{194}{41}{36}$ & $\errtwa{     0.63}{     0.06}{     0.05}$ &
 $\errtwa{     3.3} 
 {     0.1}{     0.1}$ & $\errtwa{     2.07}{     0.06}{     0.07}$
 & \nodata & $\errtwa{27.9}{0.4}{0.4}$ &  15.7/53 &    0.30 \\  
\noalign{\vspace*{0.7mm}}
02 & \nodata & \nodata & \nodata & \nodata & $\errtwa{1.96}{0.12}{0.12}$ &
 $\errtwa{3.6}{0.3}{0.3}$ & 
 $\errtwa{0.73}{0.04}{0.03}$ & $\errtwa{0.92}{0.02}{0.02}$ & 
 $\errtwa{34.0}{3}{2}$ &  67.6/80 &
 0.84 \\  
\noalign{\vspace*{0.7mm}}
03 & $\errtwa{    6.36}{    0.21}{    0.16}$ & $\errtwa
 {     0.8}{     0.2}{     0.2}$ & $\errtwa{    2.42}{    0.44}{    0.49}$
 & $\errtwa{182}{33}{37}$ & $\errtwa{     0.66}{     0.06}{     0.06}$ &
 $\errtwa{     3.4} 
 {     0.1}{     0.1}$ & $\errtwa{     1.95}{     0.06}{     0.06}$
 & \nodata & $\errtwa{27.9}{0.4}{0.4}$ &  16.7/53 &    0.31 \\  
\noalign{\vspace*{0.7mm}}
03 & \nodata & \nodata & \nodata & \nodata & $\errtwa{2.21}{0.48}{0.34}$ &
 $\errtwa{3.4}{0.7}{1.1}$ & 
 $\errtwa{0.61}{0.17}{0.08}$ & $\errtwa{1.08}{0.04}{0.04}$ & 
 $\errtwa{37.4}{26}{7}$ &  72.4/80 &
 0.90 \\  
\noalign{\vspace*{0.7mm}}
04 & $\errtwa{    6.40}{    0.22}{    0.17}$ & $\errtwa
 {     0.8}{     0.2}{     0.2}$ & $\errtwa{    1.53}{    0.32}{    0.34}$
 & $\errtwa{174}{36}{39}$ & $\errtwa{     0.70}{     0.08}{     0.07}$ &
 $\errtwa{     3.6} 
 {     0.1}{     0.1}$ & $\errtwa{     1.27}{     0.05}{     0.05}$
 & \nodata & $\errtwa{26.6}{0.4}{0.4}$ &  13.5/53 &    0.25 \\  
\noalign{\vspace*{0.7mm}}
04 & \nodata & \nodata & \nodata & \nodata & $\errtwa{2.09}{0.19}{0.22}$ &
 $\errtwa{3.4}{0.6}{0.5}$ & 
 $\errtwa{0.46}{0.04}{0.04}$ & $\errtwa{0.98}{0.02}{0.02}$ & 
 $\errtwa{36.9}{9}{6}$ &  78.1/80 &
 0.98 \\  
\noalign{\vspace*{0.7mm}}
05 & $\errtwa{    6.28}{    0.31}{    0.97}$ & $\errtwa
 {     0.9}{     0.3}{     0.3}$ & $\errtwa{    0.52}{    0.73}{    0.17}$
 & $\errtwa{138}{100}{100}$ & $\errtwa{     0.68}{     0.15}{     0.08}$ &
 $\errtwa{     4.4} 
 {     0.3}{     0.1}$ & $\errtwa{     0.53}{     0.02}{     0.02}$
 & \nodata & $\errtwa{20.6}{0.4}{0.4}$ &  20.5/53 &    0.39 \\  
\noalign{\vspace*{0.7mm}}
05 & \nodata & \nodata & \nodata & \nodata & $\errtwa{2.89}{0.78}{0.65}$ &
 $\errtwa{3.1}{1.1}{2.0}$ & 
 $\errtwa{0.16}{0.03}{0.03}$ & $\errtwa{0.99}{0.05}{0.05}$ & 
 $\errtwa{44.9}{103}{13}$ &  71.8/80 &
 0.90 \\  
\noalign{\vspace*{0.7mm}}
06 & $\errtwa{    6.26}{    0.24}{    0.22}$ & $\errtwa
 {     0.9}{     0.2}{     0.2}$ & $\errtwa{    2.06}{    0.50}{    0.46}$
 & $\errtwa{186}{45}{42}$ & $\errtwa{     0.66}{     0.07}{     0.05}$ &
 $\errtwa{     3.6} 
 {     0.1}{     0.1}$ & $\errtwa{     1.57}{     0.04}{     0.06}$
 & \nodata & $\errtwa{26.2}{0.4}{0.4}$  &  16.5/53 &    0.31 \\  
\noalign{\vspace*{0.7mm}}
06 & \nodata & \nodata & \nodata & \nodata & $\errtwa{2.33}{0.21}{0.18}$ &
 $\errtwa{3.1}{0.6}{0.4}$ & 
 $\errtwa{0.54}{0.08}{0.04}$ & $\errtwa{0.93}{0.02}{0.02}$ & 
 $\errtwa{42.3}{11}{8}$ &  88.7/80 &
 1.11 \\  
\noalign{\vspace*{0.7mm}}
07 & $\errtwa{    6.40}{    0.19}{    0.18}$ & $\errtwa
 {     0.8}{     0.2}{     0.2}$ & $\errtwa{    2.67}{    0.56}{    0.49}$
 & $\errtwa{192}{41}{35}$ & $\errtwa{     0.65}{     0.06}{     0.05}$ &
 $\errtwa{     3.3} 
 {     0.1}{     0.1}$ & $\errtwa{     2.06}{     0.05}{     0.07}$
 & \nodata & $\errtwa{28.4}{0.4}{0.4}$  &  15.6/53 &    0.30 \\  
\noalign{\vspace*{0.7mm}}
07 & \nodata & \nodata & \nodata & \nodata & $\errtwa{1.99}{0.13}{0.16}$ &
 $\errtwa{3.1}{0.4}{0.4}$ & 
 $\errtwa{0.75}{0.09}{0.04}$ & $\errtwa{0.96}{0.02}{0.02}$ & 
 $\errtwa{40.7}{10}{6}$ & 102.9/80 &
 1.29 \\  
\noalign{\vspace*{0.7mm}}
08 & $\errtwa{    6.34}{    0.23}{    0.21}$ & $\errtwa
 {     0.9}{     0.2}{     0.2}$ & $\errtwa{    1.74}{    0.40}{    0.40}$
 & $\errtwa{189}{43}{43}$ & $\errtwa{     0.71}{     0.08}{     0.07}$ &
 $\errtwa{     3.7} 
 {     0.1}{     0.1}$ & $\errtwa{     1.30}{     0.05}{     0.05}$
 & \nodata & $\errtwa{25.9}{0.4}{0.4}$ &  15.5/53 &    0.29 \\   
\noalign{\vspace*{0.7mm}}
08 & \nodata & \nodata & \nodata & \nodata & $\errtwa{2.37}{0.29}{0.23}$ &
 $\errtwa{3.7}{0.4}{0.5}$ & 
 $\errtwa{0.43}{0.03}{0.03}$ & $\errtwa{0.98}{0.02}{0.02}$ & 
 $\errtwa{34.6}{7}{5}$ & 109.0/80 &
 1.36 \\  
\noalign{\vspace*{0.7mm}}
\hline
\end{tabular}

%%% Local Variables: 
%%% mode: latex
%%% TeX-master: t
%%% End: 
}
\tablecomments{The three fit parameters of the Comptonization
  model are the compactness of the corona, $\ell_{\rm c}$, the coronal
  optical depth, $\tau_{\rm c}$, and a normalization constant, $A_{\rm
  kot}$. The Gaussian line is parameterized as in the previous tables. From
  the best fit parameters, the equivalent width of the line, EW, and the
  density averaged coronal temperature, $kT_{\rm c}$, are derived.}
\end{table*}

Attempts to fit these models to the joint PCA/HEXTE data failed.  Typical
$\chi^2_{\rm red}$ values, even allowing for the inclusion of an extra
Gaussian line component, were $\aproxgt 1.3$. These fits showed a clear
tendency for a hardening in the HEXTE band, and therefore we considered
them to be influenced by the cross-calibration uncertainties between the
PCA and the HEXTE instruments (Note that our previous fits of Cyg~X-1 used
an earlier version of the PCA response where we applied 1.5\% systematic
uncertainties across the \emph{entire} PCA band; these fits yielded
$\chi^2_{\rm red} \approx 1.6$, \emph{without} considering an additional
Gaussian component).  We therefore considered `sphere+disk' models fit to
the PCA and the HEXTE data separately. In Table~\ref{tab:jwstud3}, we
present the best-fit parameters for these models applied to our GX~339$-$4
data.

Although our numerical `sphere+disk' models do include reflection and a
fluorescent Fe line (typical equivalent width $\approx 25$\,eV) from the
disk, the PCA data showed residuals in the 5--7\,keV band,  similar
as in in our fits to RXTE data of Cyg~X--1 (\cite{dove:98a}).
We included an additional Gaussian component to our fits.  The equivalent
widths of the additional lines were $\approx 150$\,eV, and they appeared to
be broad ($\sigma \approx 0.8$\,keV).  This additional line may be
attributable partly to uncertainties in the PCA response. For these
fits, as well as for the reflection model fits, lines with energies
significantly less than 6.4\,keV can be fit, and this is likely a
systematic effect. Part of the discrepancy between the data and the model,
however, is significant. As we have discussed for our fits to
the RXTE Cyg~X--1 data, there are several possible physical interpretations
for the additional required equivalent width:  There may be an overlap
between the disk and the sphere (our models invoke a sharp transition), the
disk may be flared (we model a flat disk), the disk may have non-solar
abundances, or alternatively one might invoke a `patchy disk' embedded in
the corona (\cite{zdziarski:98a}).  The best-fit reflection fractions
of $f \approx 0.4$--$0.5$ found above are further indication that our
models may require an additional source of reflected flux.

Allowing an additional Gaussian line component, the fits to the PCA data
yield extremely low $\chi^2_{\rm red}$, which could be indicating that we
are partly fitting systematic features in the PCA response.  Note also that
the PCA data fits yielded consistently larger optical depths and
consistently lower compactness parameters than the HEXTE data fits.  The
latter was more significant, and is again indicative of the HEXTE response
being harder than the PCA response. Both instruments yielded optical depths
$\tau_{\rm c} \approx 3$--4; however, due to the discrepancy in the
best-fit spectral slopes between the PCA and the HEXTE bands, the best-fit
average coronal temperatures range from 21--30\,keV (PCA) to 34--45\,keV
(HEXTE).

\subsubsection{ADAF Models}\label{sec:adaf} 

The basic picture of mass accretion via an ADAF in the context of galactic
BHC was introduced by \citey{ichimaru:77a} and has been elaborated upon in
a series of papers by Narayan and collaborators
(\cite{narayan:97a,esin:97c}).  The accretion flow is divided into two
distinct zones: the inner part is modeled as a hot, optically thin ADAF
similar in some respects to the spherical corona discussed above, while the
outer part consists of a standard optically thick, geometrically thin disk.
The transition radius between the two zones, $r_{\rm tr}=R_{tr}/R_{\rm G}$,
is one of the model parameters.  We compute the ADAF spectrum according to
the procedure described by \citey{dimatteo:98a}.  The electrons in an ADAF
cool via three processes: bremsstrahlung, synchrotron radiation, and inverse
Compton scattering.  In addition we add the emission from a thin disk---
calculated as a standard multicolor blackbody--- and include the Compton
reflection component due to the scattering of high energy photons incident
on the disk.

In the ADAF models discussed here, we fix the black hole mass to be $m
\equiv M/M_\odot = 6$, assume the magnetic field to be in equipartition
with thermal pressure ($\beta =0.5$), and set the standard Shakura-Sunyaev
viscosity parameter (\cite{shakura:73a}) to be $\alpha_{\rm SS} =0.3$. We
normalize the accretion rate to $\dot m \equiv \dot M c^2 / L_{\rm Edd}$,
where $L_{\rm Edd}$ is the Eddington luminosity of the source.  The hard
state corresponds to mass accretion rates $\dot{m} \le \dot{m}_{\rm
  crit}=10^{-2}$, where $\dot{m}_{\rm crit}$ is the critical value above
which an ADAF no longer exists. As $\dot{m}$ increases towards
$\dot{m}_{\rm crit}$, the scattering optical depth of the ADAF goes up
which causes the spectrum to become harder and smoother. Most of the flux
from the ADAF plus disk configuration is emitted around 100\,keV and the
spectrum falls off at higher energies.

\begin{figure*}
\centerline{
\psfig{figure=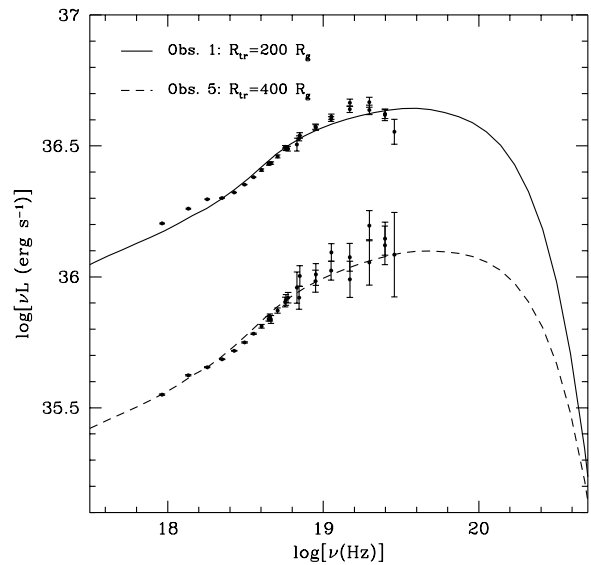,width=0.45\textwidth} }
\caption{\small Advection Dominated Accretion Flow models for the unfolded
  RXTE data from Observation 1 (solid line) and Observation 5 (dash-dot
  line).  Parameters consistent with the unfolded data are described in the
  text. A source distance of 4 kpc was assumed.}
\label{fig:tiz}
\end{figure*}

The model spectrum changes mainly as a function of $r_{\rm tr}$ and
$\dot{m}$.  The various spectral states correspond to different values of
these parameters. For example, \citey{esin:97c} attempt to explain the
initial transition from soft to hard seen in the decay of Nova Muscae by a
large change in $r_{\rm tr}$ (from $r_{\rm tr} \approx 10$ to $r_{\rm tr}
\approx 10^4$), followed by an exponential decay in $\dot{m}$ for the
subsequent evolution of this transient system.  The ASCA data of GX~339$-$4
discussed in \S\ref{sec:asca} imply that comparably large changes in
$r_{\rm tr}$ are not relevant to those observations.  Here, however,
unfolded RXTE data from Observation~1 and Observation~5, the brightest and
faintest observation respectively, can be described by ADAF models with
$r_{\rm tr} = 200$, $\dot m = 0.08$ and $r_{\rm tr} = 400$, $\dot m =
0.05$, respectively.  These model spectra and RXTE spectra for Observations
1 and 5 unfolded with a cutoff broken power law plus Gaussian line are
shown in Fig.~\ref{fig:tiz}.  In these ADAF models, the observed spectral
and luminosity changes of GX~339$-$4 are predominantly driven by changes in
the transition radius; the implied accretion rate change is substantially
smaller than the factor of 5 for the observed luminosity change.

For $r_{\rm tr} \sim 200$, which provides a rough description of
Observation~1, the disk blackbody emission peaks in the far UV/soft X-rays
and dominates over the synchrotron emission, which peaks in the optical/UV
(see also \cite{zdziarski:98a}).  The peak synchrotron emission frequency
scales as $\propto m^{-1/2}\dot{m}^{1/2}r_{\rm tr}^{5/4}$ and peaks in the
range $\nu = 10^{11}$--$10^{12}$\,Hz for supermassive black holes and $\nu
= 10^{15}$--$10^{16}$\,Hz for galactic black holes.  The spectrum below the
peak is approximately $S_\nu \propto \nu^{2}$.  The synchrotron emission
can contribute significantly to the radio emission of super-massive black
holes (although see \cite{dimatteo:98a}); however, the predicted radio flux
of GX~339$-$4 is ten orders of magnitude below the observed 7 mJy flux at
843 MHz.  Thus, there must be an extended source of radio emission, which
we further discuss in the next section.

\section{Simultaneous Radio Observations}\label{sec:radio}
The first three of our RXTE observations were simultaneous with 843\,MHz
observations taken with the Molongolo Observatory Synthesis Telescope
(MOST), and with 8.3--9.1\,GHz observations taken at the Australian
Telescope Compact Array (ATCA).  Extensive discussion of the MOST and ATCA
observations can be found in \citey{hanni:98a} and \citey{corbel:98a},
respectively (see also Table~\ref{tab:log}).

An estimate of the minimum size of the radio emitting region can be
obtained by noting that observationally the brightness temperatures of
radio sources usually are not larger than $10^{12}$\,K, else the electrons
will suffer catastrophic inverse Compton losses.  The brightness
temperature of a uniformly bright spherical source is given by $(c D/d
\nu)^2 S_\nu/2\pi k$, where $d$ is the diameter of the source, $D$ is its
distance, $\nu$ is the observed radio frequency, $S_\nu$ is the observed
radio flux density, $c$ is the speed of light, and $k$ is the Boltzmann
constant.  Taking the 7\,mJy observed at 843\,Mz by MOST and the fact that
GX~339$-$4 is unresolved, we derive
\begin{equation}
d \aproxgt 4 \times 10^{12}\,{\rm cm} \left ( {{D}\over{4\,{\rm kpc}}}
  \right ) \approx 3 \times 10^6\,R_{\rm G} \left ( {{D}\over{4\,{\rm kpc}}}
  \right ) \left  ( {{M}\over{10\,M_\odot}} \right )^{-1} ,
\label{eq:bright}
\end{equation}
which is orders of magnitude larger than the inferred size of the X-ray
emitting region, even for models that posit extremely extended coronae
(e.g., \cite{esin:97c,kazanas:97a}).

This size scale is $\propto \nu^{-1}$, so that emission at 8.6 GHz could
arise in a region an order of magnitude smaller than that responsible for
the emission at 843 MHz. Indeed, the flat spectrum emission is likely to
arise in a conical jet with a radially decreasing optical depth (e.g.,
\cite{hjellming:88a}).  Thus the outflow likely has an extent of ${\cal
  O}(10^7~GM/c^2)$ or greater.  Similar estimates for source size have been
made for the other persistent black hole candidate and Z-source neutron
star X-ray binaries by \citey{fender:99a}.

Assuming a radio spectral index of $\alpha = 0.1$ and a sharp cutoff at
$10\mu {\rm m}$ (a reasonable upper limit for where the radio flux becomes
optically thin, and typical of where ADAF models become optically thin in
the radio), the radio flux is approximately 0.1\% of the 3--100\,keV X-ray
flux.  The correlation between the X-ray and radio fluxes found by
\citey{hanni:98a}, comparable to the X-ray/radio correlation observed in
Cyg~X-1 (\cite{pooley:98a}), suggests that there is a coupling between the
inner accretion disk and the extended outflow on timescales of 7 days or
less.  Matter leaving the corona at the escape velocity (0.25$c$ at $30
R_{\rm G}$) and thereafter decelerating under the influence of gravity
would take roughly 7~days to travel a distance of $10^7 R_{\rm G}$. As 7
days is the upper bound to the correlation timescale, the radio emitting
outflow must leave at slightly greater than escape velocity, or there must
be at least some amount of acceleration of the outflow.

Although the radio observations are strictly simultaneous with our first
three RXTE observations, GX~339$-$4 exhibits less than 1\% rms variability
over the shortest timescales for which a reasonable radio flux estimate can
be made ($\aproxgt 10$ minutes).  Thus there are no strong features to
correlate between the radio and X-ray bands.

\section{Discussion}\label{sec:discuss}

\smallskip
\noindent \emph{Coronal Size and Luminosity Variation:}

The relationship between the inferred size of the corona and the magnitude
of the observed flux depends upon which spectral model we are considering.
As discussed in \S\ref{sec:adaf}, for ADAF models one can associate lower
fluxes with \emph{increased} coronal radii. A larger coronal radius implies
a lower efficiency and hence a decreased observed flux, even for constant
accretion rates.  Paper~II shows that the characteristic power spectral
density (PSD) timescale for GX~339$-$4 decreases for the lowest observed
flux (Observation 5). If one associates the PSD timescale with
characteristic disk timescales, this could be in agreement with an
increased coronal radius.  However, in paper~II we also show that the time
lags between hard and soft X-ray variability \emph{decreases} with
decreasing flux, which seems counter to a positive correlation between flux
and coronal size.

The `sphere+disk' coronal models make no assumptions about the radiative
efficiency of the accretion.  The flux can be either positively or
negatively correlated with coronal radius, depending upon the variations of
the coronal compactness, $\ell_{\rm c}$, and the temperature, $T_{\rm d}$,
at the inner edge of the accretion disk that surrounds the corona
(\cite{dove:97b}).  Note that the `sphere+disk' models used in this work,
contrary to many ADAF models, do not consider synchrotron photons as a
source of seed photons for Comptonization.

Using the definitions of $\ell_{\rm c}$, $\ell_{\rm d}$, and $f$ given in
\S\ref{sec:corona}, energy balance in the `sphere+disk' system determines
the coronal radius, to within factors of order unity, to be given by
\begin{eqnarray}
R_{\rm C} &\approx& 160 
  \left ( \frac{\ell_{\rm d}+f \ell_{\rm c}}{\ell_{\rm d} + \ell_{\rm c}}
  \right )^{1/2}   
  \left ( \frac{k T_{\rm d}}{150~{\rm eV}} \right )^{-2}  
  \left ( \frac{6~M_\odot}{M} \right ) 
  \cr
  &\times& 
  \left ( \frac{D}{4~{\rm kpc}} \right )
  \left ( \frac{F_{\rm tot}}{10^{-8}\flux} \right )^{1/2} 
  R_{\rm G} ~~,
\label{eq:coronarad}
\end{eqnarray}
where $M$ is the mass of the compact object, $D$ is the distance to the
source, and $F_{\rm tot}$ is the bolometric flux of the source.  If $T_{\rm
  d}$, $f$, $\ell_{\rm d}$, and $\ell_{\rm c}$ were held fixed, then the
coronal radius would be positively correlated with flux.  Whereas this
might pose some problems for understanding the flux dependence of the
characteristic timescales observed in the PSD, this would agree with the
flux dependence of the X-ray variability time lags (paper~II).  However, as
the RXTE bandpass does not usefully extend below $\approx 3$\,keV, we do
not have a good handle on the flux dependence of $T_{\rm d}$.  If $T_{\rm
  d} \propto F_{\rm tot}^\beta$ with $\beta > 1/4$, then increasing flux
could imply decreasing coronal radius.

\smallskip
\noindent \emph{Correlations Among Spectral Parameters:} 

\begin{figure*}
\centerline{
\psfig{figure=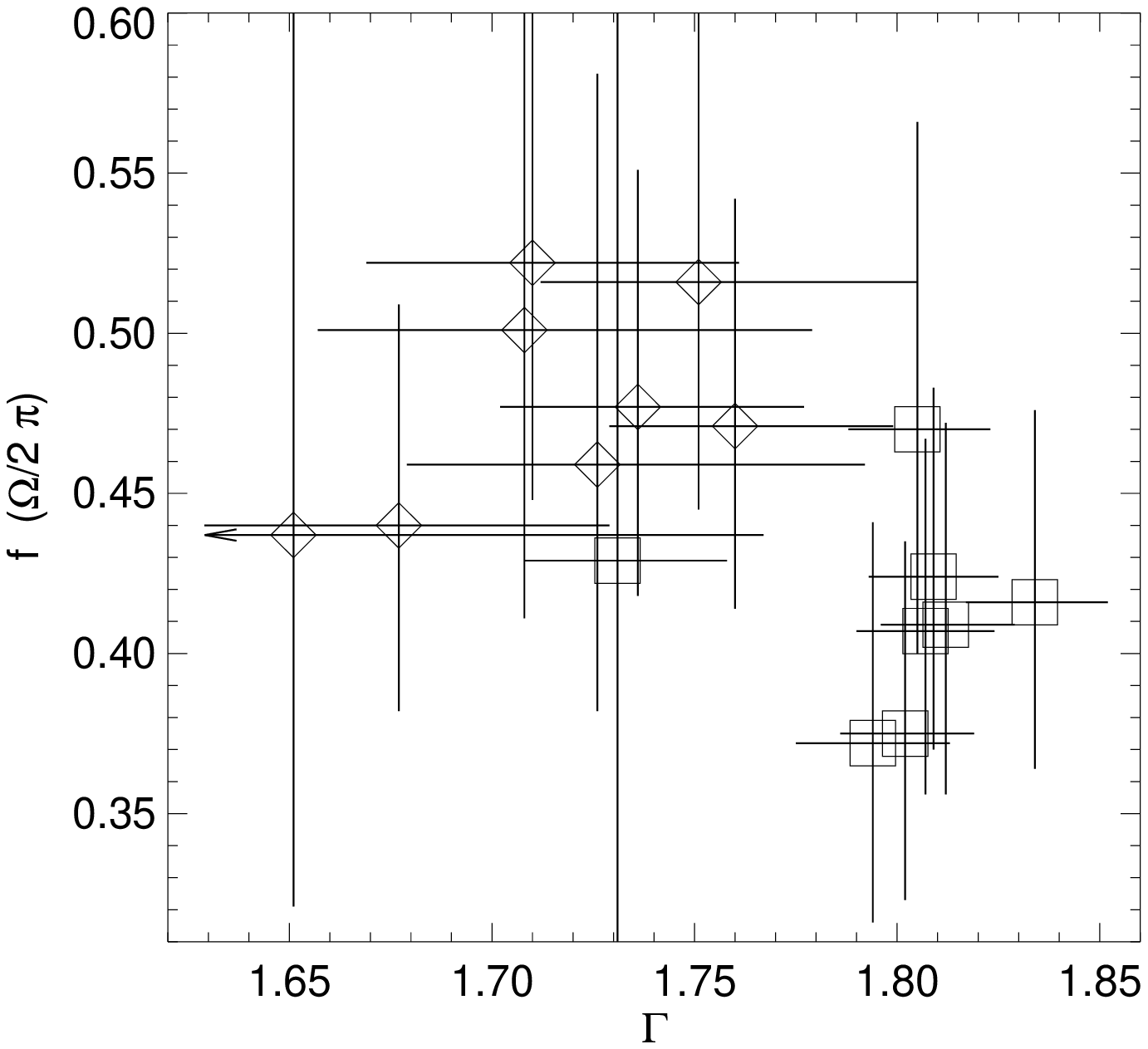,width=0.33\textwidth} 
\psfig{figure=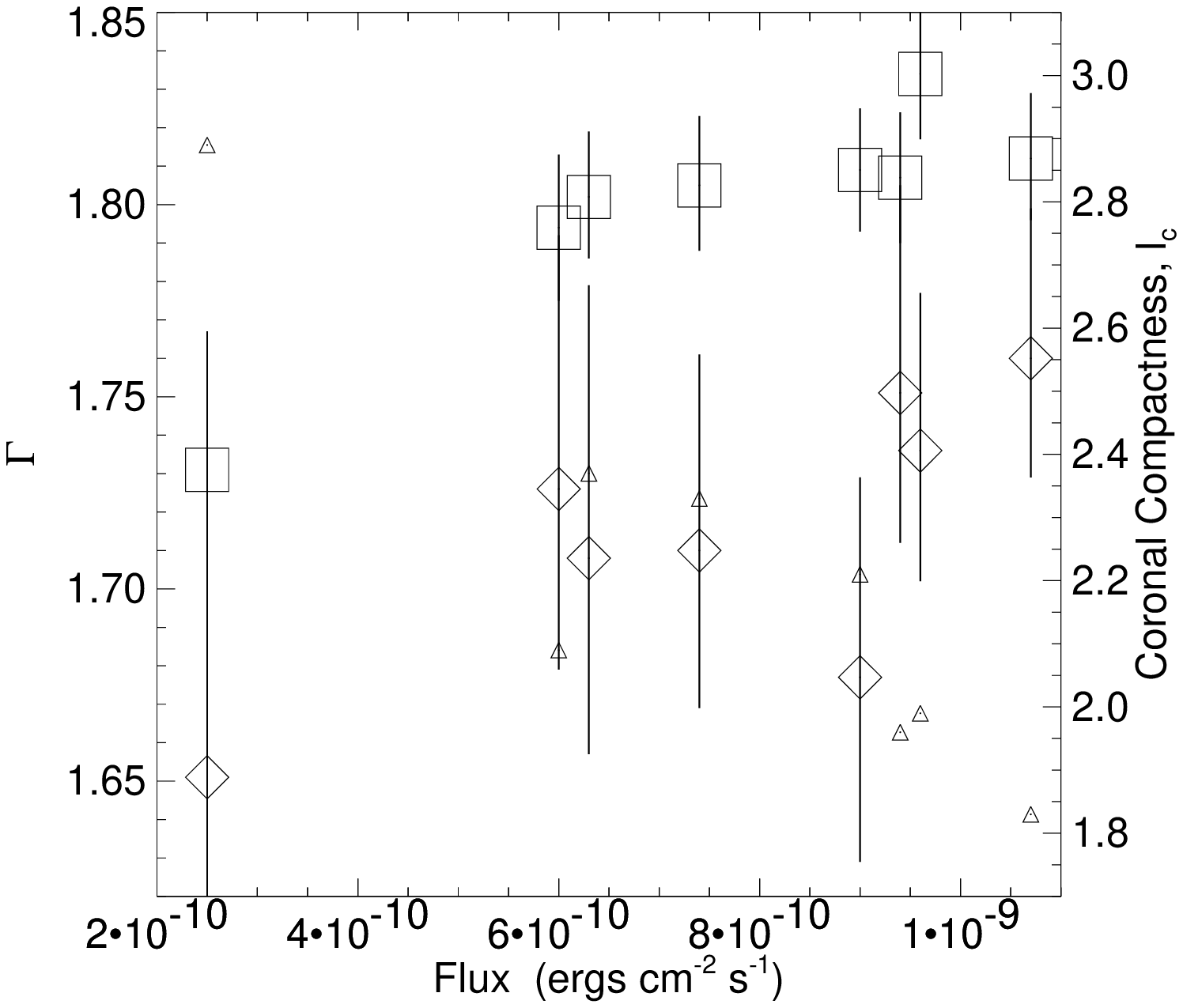,width=0.33\textwidth} 
\psfig{figure=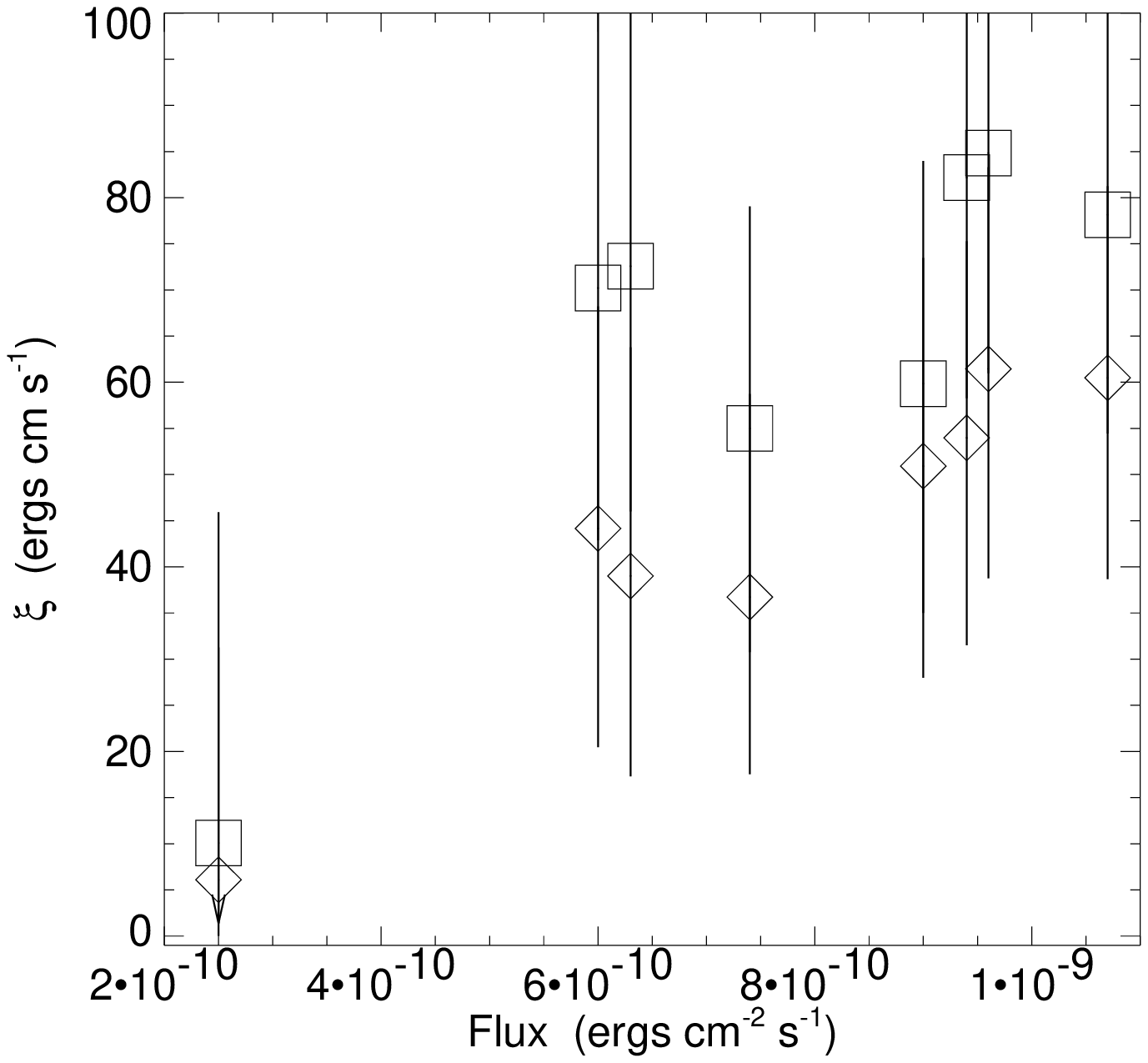,width=0.33\textwidth} 
}
\caption{\small {\it Left:} Reflection fraction vs. photon
  index, $\Gamma$, for models fit to PCA data only (squares), and models
  fit to PCA plus HEXTE data that allowed the PCA and HEXTE photon indeces
  and normalizations to be different (diamonds; HEXTE photon index shown).
  {\it Middle:} Photon index, $\Gamma$, vs. observed 3--9\,keV flux for the
  same reflection models as on the left. Also shown, without error bars, is
  the best fit compactness, $\ell_c$, for `sphere+disk' coronal models fit
  to HEXTE data (small triangles). {\it Right:} Disk ionization
  parameter, $\xi$, in units of ${\rm ergs~cm~s^{-1}}$ for the same
  reflection models as on the left. \label{fig:reflect}}
\end{figure*}

\citey{ueda:94a} claimed that reflection models of GX~339$-$4 exhibited a
correlation between photon index, $\Gamma$, and reflection fraction, $f$,
with softer spectra implying greater reflection.  \citey{zdziarski:98b} has
claimed that this correlation extends to reflection models of Seyfert~1
galaxies as well.  Such a correlation is not unreasonable to expect.  For
example, if we allow the corona and disk to overlap to some extent in the
`sphere+disk' model (\cite{poutanen:97b}), then we expect the increase in
the flux of seed photons to cool the corona and lead to a softer spectral
index.  Likewise, the covering fraction of the disk would be increased, in
agreement with the suggested correlation.  In Figure~\ref{fig:reflect} we
plot $f$ vs. $\Gamma$ for our reflection model fits to GX~339$-$4.
Contrary to the claims of \citey{ueda:94a} and \citey{zdziarski:98b},
however, there is no strong evidence for a correlation. Fitting the
reflection fraction with a function linear in $\Gamma$, as opposed to
fitting with the mean value of $f$, improves the $\chi^2$ of the fits by
$0.2$, which is not significant.  Fitting with the mean gives
$\chi^2_{\rm red} = 0.2$.

We do note two possible trends from the reflection model fits. First, as
has been noted for other hard state galactic black hole candidates
(\cite{tanaka:95a} and references therein), there may be a correlation
between flux and photon index with lower flux implying a harder source.
(The significance of the correlation is driven by Observation 5, the
faintest and hardest of the observations.  However, a similar correlation
is also present in color-intensity diagrams.)  Such a correlation is
consistent with the expectations of ADAF models where the radius increases
with decreasing flux (Figure~\ref{fig:tiz}). Again, the `sphere+disk'
corona models do not predict a clear trend without knowing the flux
dependences of other parameters such as $T_{\rm d}$.

Second, the ionization parameter, $\xi$, is positively correlated with
flux.  Such a correlation was noted by \citey{zycki:98a} for Ginga
observations of Nova Muscae.  It is not unreasonable to expect the disk to
become increasingly ionized with increasing flux.  We again caution that it
is dangerous to make one-to-one correlations between a model fit parameter
and a true physical parameter.  Furthermore, the significance of the
correlation is again almost entirely determined by Observation 5, the
faintest observation, which has $\xi \approx 0$. However, if we take the
flux dependence of $\xi$ as being real and interpret it physically, it
provides some constraints on the flux dependence of the coronal radius.
The ionization parameter is $\propto F_{\rm tot} / (\rho R^2)$, where
$\rho$ is the density of the disk.  For a gas pressure-dominated
Shakura-Sunyaev $\alpha$-disk, $\rho \propto R^{-1.65}$
(\cite{shakura:73a}).  In order for $\xi$ to be roughly linear in flux (the
actual dependence is not strongly constrained by the data), we require
$F_{\rm tot} \proptwid R^{1.35}$.  (For the `sphere+disk' models this would
further require $T_{\rm d} \proptwid R^{-1/3}$, depending upon the flux
dependences of $\ell_{\rm d}$, $\ell_{\rm c}$, $g$, etc.) Taken physically
and in the context of a gas pressure dominated Shakura-Sunyaev
$\alpha$-model, the flux dependence of $\xi$ implies that the coronal
radius increases with increasing flux.

\section{Summary}\label{sec:summ}

We have presented a series of observations of the black hole candidate
GX~339$-$4 in low luminosity, spectrally hard states.  These observations
consisted of three separate archival ASCA and eight separate RXTE data
sets.  All of these observations exhibited (3--9\,keV) flux $\aproxlt
10^{-9}~{\rm ergs~s^{-1}~cm^{-2}}$, and the observed fluxes spanned roughly
a factor of 5 in range for both the ASCA and RXTE data sets.  Subject to
uncertainties in the cross calibration between ASCA and RXTE, the faintest
ASCA observation was approximately a factor of two fainter than the
faintest RXTE observation.

All of these observations showed evidence for an $\approx 6.4$\,keV Fe line
with equivalent widths in the range of $\approx 20$--$140$\,eV.  The ASCA
observations further showed evidence for a soft excess that was
well-modeled by a power law plus a multicolor blackbody spectrum with peak
temperatures in the range $\approx 150-200$\,eV.  Both of these factors
considered together argue against `sphere+disk' or ADAF type-geometry
coronae with extremely large coronal radii of ${\cal O}(10^4~R_{\rm G})$
(e.g., \cite{esin:97c}).

The RXTE data sets were well-fit by `sphere+disk' Comptonization models
with coronal temperatures in the range $20$--$50$\,keV and optical depths
in the range of $\tau \approx 3$.  These fits were similar to our previous
fits to RXTE data of Cyg~X-1. Advection Dominated Accretion Flow models,
which posit a similar geometry, also provided reasonable descriptions of
the unfolded RXTE data.  The `sphere+disk' and ADAF models were not able,
however, to also model the observed radio fluxes. Thus, a static corona
seems to be ruled out by the observations.  The ADAF models can imply that
the coronal radius increases with decreasing flux.  The `sphere+disk'
corona models do not make a specific prediction for the dependence of the
coronal radius on the flux; however, they can be consistent with a positive
correlation between coronal radius and flux.  As described in paper~II, a
positive correlation between flux and coronal radius is consistent with the
observed flux dependence of the time lags between hard and soft X-ray
variability.

We also considered `reflection models' of the RXTE data.  These models
showed evidence of a hardening of the RXTE spectra with decreasing X-ray
flux. They further showed evidence of a positive correlation between the
best-fit ionization parameter, $\xi$, and the observed flux.  Especially
the latter of these correlations, however, was dominated by the model fits
of the faintest observation. The reflection models did not exhibit any
evidence of a correlation between the photon index of the incident power
law flux and the solid angle subtended by the reflector.

Three of the RXTE observations were strictly simultaneous with 843\,MHz and
8.3--9.1\,GHz radio observations.  The most likely source of the radio flux
is synchrotron emission from an extended outflow with a size of ${\cal
  O}(10^7~GM/c^2)$.  The correlation between radio and X-ray emission on
timescales of 7 days or less (\cite{hanni:98a}) implies a strong coupling
of the inner disk accretion flow with this spatially extended outflow as is
expected by recent theoretical arguments (\cite{blandford:98a}).  Further
simultaneous radio/X-ray observations, preferably with the addition of
IR/optical monitoring to constrain the location of the synchrotron break
and with the addition of soft X-ray monitoring to constrain the accretion
disk parameters, are required to test such models in detail.

\acknowledgements We would like to thank Dr. Christopher Reynolds for
keeping a stiff upper lip while explaining ASCA data analysis to us.  We
would also like to thank K.~Mukai of the ASCA GOF for useful advice.
W.A.~Heindl and D.~Gruber kindly provided assistance with the HEXTE data
extraction, and S.~Corbel provided assistance with the radio data.  We are
grateful to B.~Stern for writing the original version of kotelp, and, more
importantly, for finally telling us what the name means (`cauldron').  We
would also like to acknowledge useful conversations with M.~Begelman, J.
Chiang, B.A.~Harmon, K.~Pottschmidt, R.~Staubert, C.~Thompson, and
A.~Zdziarski.  This work has been financed by NASA Grants {NAG5-4731} and
{NAG5-3225} (MAN, JBD). MN was supported in part by the National Science
Foundation under Grant No. Phy94-07194. JW was supported by a travel grant
from the Deutscher Akademischer Austauschdienst, RPF was supported by an EC
Marie Curie Fellowship (ERBFMBICT 972436), and TDM thanks Trinity College
and PPARC for financial support.  This research has made use of data
obtained through the High Energy Astrophysics Science Archive Research
Center Online Service, provided by the NASA/Goddard Space Flight Center.

\appendix

\section{ASCA Data Extraction}\label{sec:ascaapp}

We extracted data from the two solid state detectors (SIS0, SIS1) and the
two gas detectors (GIS2, GIS3) onboard ASCA by using the standard ftools as
described in the ASCA Data Reduction Guide (\cite{day:98a}).  The data
extraction radius was limited by the fact that all the observations were in
1-CCD mode and that the source was typically placed close to the chip edge.
We chose circular extraction regions with radii of $\approx 4$\,arcmin for
SIS0, $\approx 3$\,arcmin for SIS1 (the maximum possible extraction radii
for these detectors), and $\approx 6$\,arcmin for GIS2 and GIS3. For
observation~3 we excluded the central 40\,arcsec to avoid the possibility
of photon pileup.  We used the sisclean and gisclean tools (with default
values) to remove hot and flickering pixels. We filtered the data with the
strict cleaning criteria outlined in the ASCA Data Analysis Handbook;
however, we took the larger value of 7\,$\mbox{GeV}/c$ for the rigidity.
We rebinned the spectral files so that each energy bin contained a minimum
of 20\,photons. We retained SIS data in the 0.5 to 9\,keV range and GIS
data in the 1.5 to 9\,keV range.

We accounted for the cross-calibration uncertainties of the three
instruments by introducing a multiplicative constant for each detector in
all of our fits.  Relative to SIS0, the SIS1 detector normalization was
always found to be within 2\%, the GIS2 normalization was found to to be
within 9\%, and the GIS3 normalization was found to be within 15\%.  For
any given observation, the normalization constants varied by $\aproxlt \pm
1\%$ for different spectral fits.  The background was measured from
rectangular regions on the two edges of the chip farthest from the source
(SIS data), or from annuli with inner radii $>6$\,arcmin (GIS data).  These
data were cleaned and filtered in the same manner as the source files.

The resulting data files showed reasonable agreement between all four
detectors.  The most discrepant detector was SIS1, which also was the
detector limited to the smallest extraction radius.  This detector tended
to show deviations from the other detectors for energies $\aproxgt 9$\,keV,
and from the SIS0 detector for energies $\approx 0.5$--$1$\,keV.  The
detectors were mostly in mutual agreement for the lowest flux observations.
It is likely that the agreement could be further improved for observations
located closer to the center of the chips (thereby allowing larger
extraction radii), and if low galactic latitude dark sky observations in
1-CCD mode were available to use as background.

\section{RXTE Data Extraction}\label{sec:rxteapp}
In \S\ref{sec:xte} we present data from both pointed instruments on RXTE,
the Proportional Counter Array, \pca, and the High Energy X-ray Timing
Experiment, \hexte. As we showed in \S\ref{sec:xte}, the large effective
area of these instruments results in a data analysis approach that is
dominated by the calibration uncertainty of these detectors (especially the
PCA). In this appendix we summarize the major properties of both
instruments and study their (inter-)calibration. All RXTE results obtained
in this paper were obtained using the standard RXTE data analysis software,
ftools version 4.1 (including the RXTE-patch 4.1.1 and the correct
accounting of the time-dependence of the PCA response; Jahoda, 1998, priv.\ 
comm.).  Spectral modeling was done using XSPEC, version 10.00s
(\cite{arnaud:96a}).

\begin{figure*}
\centerline{
\psfig{figure=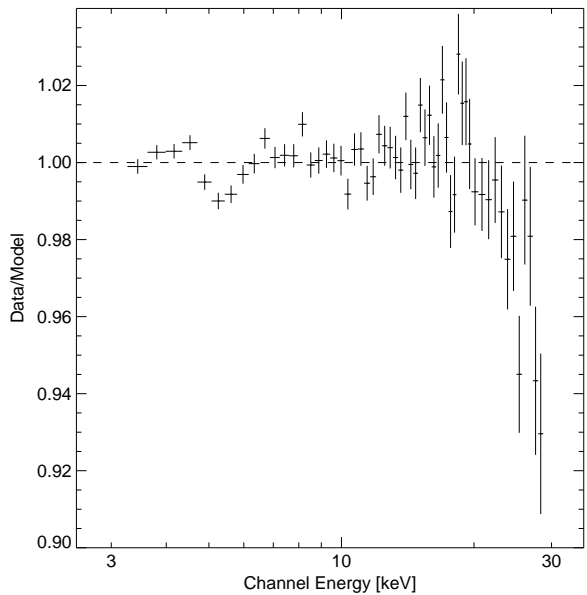,width=0.45\textwidth}}
\caption{\small Ratio between the power-law fit of eq.~(\ref{eq:crabpca})
  and the \pca\ Crab nebula data. The major deviations between the data and
  the model are due to the Xe $\rm L_{1,2,3}$ edges at $\sim 5.1$\,keV, and
  due to the Xe K edge at 34.6\,keV.}\label{fig:crabrat}
\end{figure*}

The \pca\ consists of five nearly identical co-aligned Xenon proportional
counter units (PCUs) with a total effective area of about 6500\,cm$^{2}$.
The instrument is sensitive in the energy range from 2\,keV to $\sim
60$\,keV (\cite{jahoda:96b}).  We only used data where all five PCUs were
turned on.  Background subtraction was done in the same manner as for our
RXTE Cyg~X-1 observations (\cite{dove:98a}). Specifically, a model using
the rate of Very Large Events in the detector was used to estimate the
background flux.  The major uncertainty of this estimated background is due
to activation of radioactivity in the detectors during SAA passages. Since
this background component is present for about 30~minutes after the
passage, we ignored data measured during these intervals. Furthermore, data
were not accumulated at times of high electron contamination. The electron
contamination is measured by a certain ratio of veto rates in the
detectors, the so-called ``electron ratio'' (Jahoda et al., 1998, in
preparation). As recommended by the RXTE Guest Observer's Facility (GOF),
we excluded times during which the ``electron ratio'' was larger than 0.1
in at least one of the detectors.  Note that the observed count rates from
GX 339$-$4 are too high to allow the use of the newer background model made
available by the RXTE GOF in 1998 June.

For spectral fitting, we limited the energy range of the PCA data from 3 to
30\,keV and used version 2.2.1 of the \pca\ response matrices.  These
matrices are newer than those used by us previously (\cite{dove:98a}), and
they are primarily characterized by assuming a higher instrumental
resolution (Jahoda et al., 1998, in preparation). Due to the large PCA
count rate of GX~339$-$4 ($\sim 800$\,cps) our observations are dominated
by the remaining uncertainties in the detector calibration, and not by
Poisson errors. Therefore, a good understanding of these uncertainties is
necessary.

Since the Crab spectrum is commonly assumed to be a featureless power-law
(\cite{toor:74a}), at least over narrow energy ranges, the ratio between
the fit to the Crab and the data can be used to deduce the systematic
uncertainty associated with the detector calibration.  We therefore
extracted a public domain spectrum of the Crab nebula and pulsar measured
with the PCA on 1997 April~1. The Crab data were screened using the same
criteria as those applied to our GX~339$-$4 data, except for that the
``electron ratio'' check was not applied since the background contributes
only 0.6\% to the total number of photons detected for the Crab.  Modeling
the 3 to 30\,keV Crab data with an absorbed power-law resulted in a
best-fit photon spectrum of the form
\begin{equation}\label{eq:crabpca}
N_{\rm ph} = 13.3 E^{-2.187} \exp{\left[ - 2.54\times 10^{21}\,{\rm
    cm}^{-2}\sigma_{\rm bf}(E) \right]}\,{\rm
    cm^{-2}\,s^{-1}\,keV^{-1}}
\end{equation}
where $\sigma_{\rm bf}(E)$ is the energy dependent bound free absorption
cross section for material of cosmic abundances as given by
\citey{morrison:83a}, and where $E$ is the photon-energy measured in keV.
For this fit to the Crab data, $\chi^2/{\rm dof}=168/56$. In
Fig.~\ref{fig:crabrat} we display the ratio between the best-fit to the
Crab and the data. Using this ratio plot, we deduced the systematic
uncertainties of the detector (Table~\ref{tab:syst}). Adding them in
quadrature to the Poisson errors of our data, the $\chi^2$ of the above
Crab fit was reduced to $\chi^2_{\rm red}=0.56$.  Note that the internal
consistency of the detector calibration of the PCA appears to be on the 1\%
level, i.e., comparable to that obtained for previously flown instruments
like the \textsl{Ginga} LAC (\cite{turner:89b}) or the ASCA GIS
(\cite{makishima:96a}), even though the much larger effective area of the
PCA makes it necessary to include many ``dirt effects'' into the detector
model.

We caution that the statistical uncertainties of the fit parameters
presented in this work were derived using the above systematic
uncertainties. It is questionable, therefore, whether the approach of
\citey{lampton:76a} to determine the uncertainties from the
$\chi^2$-contours can really be used since this approach makes use of the
assumption of Poisson-type errors.  The uncertainties given in this work
should be taken with these caveats in mind, especially for those fits where
the $\chi^2_{\rm red}$ values are very small (i.e., $\aproxlt 0.5$), and
thus they should not be construed literally as `90\% uncertainties'.

\begin{table*}
\small{
\caption{\small Systematic uncertainties of version~2.2.1 of the \pca\ response
  matrix as obtained from a power-law fit to the Crab spectrum
  (Fig.~\ref{fig:crabrat}).}\label{tab:syst}
\smallskip
\centerline{
\begin{tabular}{llllll}
\hline
\hline
\noalign{\vspace*{1mm}}
Channel${}^{\rm a}$ & 0--15 & 16--39 & 40--57 & 58--128 \\
\noalign{\vspace*{1mm}}
Channel-Energy [keV]& 0--7  & 7--16  & 16--25 & 25--   \\
\noalign{\vspace*{1mm}}
Uncertainty [\%]    & 1     & 0.5    & 2      & 5      \\
\noalign{\vspace*{1mm}}
\hline
 \end{tabular}}
 \smallskip
 ${}^{\rm a}$\ PCA channels assuming the standard2 channel binning.}
 \end{table*}

 The \hexte\ consists of two clusters of four NaI/CsI-phoswich
 scintillation counters, sensitive from 15 to 250\,keV.  A full description
 of the instrument is given by \citey{rothschild:98a}.  Source-background
 rocking of the two clusters provides a direct measurement of the \hexte\ 
 background with measured long term systematic uncertainties of $<1\%$
 (\cite{rothschild:98a}).  Although no other strong sources in the field
 around GX~339$-$4 are known
 (\cite{covault:92a,bouchet:93a,trudolyubov:98a}), we extracted individual
 background spectra for both \hexte\ cluster background positions to check
 for contamination of the spectrum from weak background sources.  In all
 cases the background spectra differed by $\lesssim 1$\,cps. Thus, we used
 the added background spectra from both cluster positions in our data
 analysis.  We used the standard response matrices dated 1997~March~20 and
 used data measured between 17 and 110\,keV.  An analysis of the detector
 calibration similar to that performed for the \pca\ reveals that the
 \hexte\ calibration is good on a level comparable to the \pca. Due to the
 much smaller effective area of the detector and due to the smaller flux
 from the source at higher energies, however, the \hexte\ spectrum is
 completely dominated by the Poisson error of the data. Therefore we did
 not attempt to take the systematic calibration uncertainty into account.
 To improve the statistics of individual energy bins, we rebinned the raw
 ($\approx1$\,keV wide) energy channels by a factor of 2 for the energy
 range from 30 to 51\,keV, and by a factor of 3 above that.

 When modeling the spectrum of GX~339$-$4 from 3 to 110\,keV as measured by
 the \pca\ and the \hexte, the intercalibration between the instruments is
 of some concern. Our experience with previous data and the response matrix
 described by \citey{dove:98a} indicated that the flux calibration of the
 \hexte\ with respect to the \pca\ was off by about 25\%, i.e., the derived
 \hexte\ fluxes were $\sim75$\% of those found with the \pca. This deviation
 is mainly due to a slight misalignment of the \hexte\ collimators which has
 not yet been taken into account in the \hexte\ response matrix (Heindl
 1998, priv.\ comm.). Using the new \pca\ response, the flux ratio now
 appears to be larger and was found to be $62$---$69\%$ in our data.
 Extracting spectra with internal software used by the HEXTE hardware team
 produced spectra identical to those found using our extraction procedure.
 Therefore we do not believe this change in the flux calibration to be due
 to errors in the deadtime correction.  To take this offset in the effective
 areas into account we modeled the spectra using a multiplicative constant
 which was set to unity for the \pca, and which was a fit parameter for both
 \hexte\ clusters.  Thus, all fluxes given below were measured with respect
 to the PCA. The maximum deviation of the HEXTE clusters relative to each
 other was found to be less than 8\%.

 Apart from the flux calibration, however, even more crucial for our
 analysis is the question of how well the inferred spectral shapes agree
 for the two instruments. Our Crab fits show that the \pca\ results in a
 photon index of $\Gamma=2.187$ (Eq.~\ref{eq:crabpca}), while our \hexte\ 
 fits gave $\Gamma=2.053$. The generally accepted value for the Crab photon
 index in the 1--100\,keV range is $\Gamma=2.10\pm 0.03$. This value was
 adopted by \citey{toor:74a} in their analysis of 28 different rocket
 flight measurements.  There are indications that the spectrum softens to
 $\Gamma\sim 2.5$ above 150\,keV (\cite{jung:89a}). Although the absolute
 uncertainty of the Crab flux in the 2--100\,keV range has been estimated
 as large as 2\%, and even larger below 2\,keV
 (\cite{noergaard-nielsen:94a}), the $\Delta \Gamma = 0.134$ deviation
 between the \pca\ and the \hexte\ best-fit Crab photon index is still very
 worrisome, and it is currently being studied by both the PCA and HEXTE
 instrument teams (Jahoda 1998, priv. comm.).

 %\bibliographystyle{jwapjbib}
 %\bibliography{mnemonic,jw_abbrv,apj_abbrv,bhc,agn,diplom,inst,ns,conferences}

\end{document}